\documentclass[preprint,12pt]{elsarticle}

\usepackage{lineno,hyperref}
\usepackage{bm}
\usepackage{amsfonts}
\usepackage{amssymb}
\usepackage{amsmath}
\usepackage{isomath}

\usepackage{scalerel,stackengine}
\stackMath
\newcommand\reallywidehat[1]{%
\savestack{\tmpbox}{\stretchto{%
  \scaleto{%
    \scalerel*[\widthof{\ensuremath{#1}}]{\kern-.6pt\bigwedge\kern-.6pt}%
    {\rule[-\textheight/2]{1ex}{\textheight}}
  }{\textheight}%
}{0.5ex}}%
\stackon[1pt]{#1}{\tmpbox}%
}

\usepackage{stmaryrd} 
\usepackage[usenames]{color}
\usepackage{epsfig}
\usepackage{graphicx}
\usepackage{psfrag}

\graphicspath{{Figures/}}
\usepackage{float}
\floatstyle{boxed}
\newfloat{algorithm}{htb}{loa}
\floatname{algorithm}{Algorithm}
\usepackage{algorithm,algpseudocode}
\usepackage{caption}
\usepackage{subcaption}
\usepackage{color}
\usepackage{xcolor}
\usepackage{multirow}
\usepackage{natbib}
\modulolinenumbers[5]

\usepackage[left = 2cm, right=2.5cm, top=2cm, bottom =2cm]{geometry}

\begin{document}
\begin{frontmatter}





\title{A mixed formulation for physics-informed neural networks as a potential solver for engineering problems in heterogeneous domains: \\comparison with finite element method}

\author{ Shahed Rezaei$^{1}$, Ali Harandi$^{2}$, Ahmad Moeineddin$^{2}$, Bai-Xiang Xu$^{1}$, Stefanie Reese$^2$} 
\address{$^1$Mechanics of Functional Materials Division, Institute of Materials Science,\\ Technical University of Darmstadt, Otto-Berndt-Str. 3, D-64287 Darmstadt, Germany}
\address{$^2$Institute of Applied Mechanics \\ RWTH Aachen University, Mies-van-der-Rohe-Str. 1, D-52074 Aachen, Germany}
\corref{ali.harandi@ifam.rwth-aachen.de}

\begin{abstract}
Physics informed neural networks (PINNs) are capable of finding the solution for a given boundary value problem. We employ several ideas from the finite element method (FEM) to enhance the performance of existing PINNs in engineering problems. The main contribution of the current work is to promote using the spatial gradient of the primary variable as an output from separated neural networks. Later on, the strong form which has a higher order of derivatives is applied to the spatial gradients of the primary variable as the physical constraint. In addition, the so-called energy form of the problem is applied to the primary variable as an additional constraint for training. The proposed approach only required up to first-order derivatives to construct the physical loss functions. We discuss why this point is beneficial through various comparisons between different models. The mixed formulation-based PINNs and FE methods share some similarities. While the former minimizes the PDE and its energy form at given collocation points utilizing a complex nonlinear interpolation through a neural network, the latter does the same at element nodes with the help of shape functions. We focus on heterogeneous solids to show the capability of deep learning for predicting the solution in a complex environment under different boundary conditions. The performance of the proposed PINN model is checked against the solution from FEM on two prototype problems: elasticity and the Poisson equation (steady-state diffusion problem). We concluded that by properly designing the network architecture in PINN, the deep learning model has the potential to solve the unknowns in a heterogeneous domain without any available initial data from other sources. Finally, discussions are provided on the combination of PINN and FEM for a fast and accurate design of composite materials in future developments. \\

\end{abstract} 
\begin{keyword} 
Physical informed neural networks, Unsupervised learning, heterogeneous solids
\end{keyword}

\end{frontmatter}

\section{Introduction} 
Deep learning (DL) methods as a branch of machine learning (ML) algorithms and artificial intelligence have gained attention in various fields of science. The ability of these models to realize the patterns in the available data makes them very attractive candidates for fast and reliable predictions. These methods also find their way to many engineering applications, especially in material mechanics and design \cite{Bock2019}. 
One significant advantage of DL algorithms is the huge speedup that we can gain upon successful training. The latter point is particularly important when it comes to multiscaling analysis where we would like to have the material properties at the upper scale which are under the influence of the features at the lower scale. Transforming this information is a tedious task; therefore, DL can play a huge role in reducing the complexity (e.g. see \citet{Peng2021, fernandez2020, Yin2022}).

Here, we would like to examine the potential of the DL method in particular, for the field of computational engineering, where we are dealing with a complex and perhaps coupled system of governing equations. Moreover, it is vital to clarify different ways based on which one can apply machine learning methods.
The first main-stream goes toward application of DL to find the solution using supervised learning based on available data.
Clearly, in these contributions, one requires to perform enough off-line calculations or measurements to provide the required data for the network training. The performance of the models in this case theoretically should be improved with increasing training data.
\citet{HSU2020197} presented an ML approach to predict fracture processes based on data from the molecular simulation and the image-processing approaches.
Mianroodi et al. \cite{mianroodi2021teaching} applied a U-Net approach to predict stress fields in heterogeneous non-linear elastoplastic material systems.
\citet{Bhaduri2021} considered a fiber-reinforced matrix composite material system and used a convolutional neural network (CNN/U-Net architecture) to evaluate stress fields. See also \cite{mianroodi2022lossless, LIN2021109193} for finding structure-property relation utilizing CNN approaches. 
The application of DL for multiscaling is well-established and has shown its performance in various fields. See for example \cite{Peng2021, LIN2021109193, Kumar2020, Mozaffar2019, WANG2018337}. The idea here is to gather data by performing simulations utilizing well-established solvers (such as finite elements, molecular dynamics, etc.). Based on the obtained data, a proper NN is trained to transfer the information to the upper scale \cite{fernandez2020, mianroodi2022lossless}. In other words, the NN relates the inputs at the lower scale to the required outputs at the upper scale. These NNs require enough data for an accurate prediction, and they do not necessarily satisfy physical constraints governing the given problem. Some recent developments tend to enhance the architecture and design of the NNs based on the application-dependent physical constraints. See for example \cite{linka2020constitutive, MASI2021104277, Fernandez2021}. Further improvements and studies in this direction are essential. Therefore, utilizing physics-informed NN can be very advantageous for this purpose. Note that in this category, the initial and labeled data comes from other resources (either other numerical simulations or experimental measurements) and we still need developments in those fields to gather enough data for training. In other words, here the classical solvers and measures go hand in hand with the machine learning approaches.

Another mainstream is to employ DL methods to find the solution for a boundary value problem (BVP) without any labeled data and solely based on governing equations/physics (unsupervised learning). This is done by introducing the governing equations together with initial and boundary conditions as additional terms in the loss function of the neural networks. The latter is also known in the community as physics informed neural networks (PINNs) \cite{RAISSI2019}.
Here, the performance of the DL does not improve with the number of BVPs, just like a finite element solver. Early investigation in this direction seem promising \cite{RAISSI2019, Guo2022, Henkes2022}. 
The main advantage of these approaches is that the physical laws are integrated into the network's loss function. As a result, upon proper training, one is confident that the network's outcome respects physical laws even for unseen data and blind predictions in the domain of training.
In other words, one deals with a constraint minimization problem which results in the solution of the given boundary value problem \cite{RAISSI2019}. This approach is explored in earlier times to some extent. \citet{LEE1990} employed neural minimization to solve the finite difference PDEs. \citet{Psichogios1992} used the mixed formulation based neural network which utilized the available prior knowledge about the process and showed significantly less requirement of the ground truth data. 
\citet{Lagaris1998} utilized artificial neural networks for solving boundary value problems, and they compared their results to the ones in which FEM is used. The PINN is further developed, and discussed in detail for various types of PDEs in different branches of physics and engineering \citet{RAISSI2019}. It is shown how easily one has access to derivatives of the output variable with respect to the input parameters thanks to automatic differentiation (AD).
This idea was picked up by many authors and applied to various fields from fluid mechanics to solid mechanics. For a review of the approach and some recent developments, see \cite{Cai2022, karniadakis2021physics, Bock2019}. PINNs is also applied to arbitrary complex domains.
\citet{BERG201828} used deep feedforward artificial neural networks to approximate solutions to PDEs in complex geometries.
\citet{Blechschmidt2021} introduced several ways for solving PDEs via ML-based algorithms and compared the formulation of PINNs against other available approaches. 
\citet{ZOBEIRY2021104232} developed a PINN to solve conductive heat transfer equation as well as convective heat transfer in manufacturing applications.

DL methods are also applied in the context of solid mechanics. \citet{SAMANIEGO2020112790} explored deep neural networks for approximation of the solution to PDEs and analyzed the energetic format of the PDE. The energy of the given system is introduced as a natural loss function for a machine learning method, see also \cite{Guo2020}. \citet{Guo2022} presented a deep collocation method (DCM) in non-homogeneous media which utilizes a PINN with material transfer learning. The authors studied potential problems such as heat transfer and electric conduction. See also \cite{Cai2021} where authors presented applications of PINNs to heat transfer problems.
\citet{HAGHIGHAT2021} applied PINNs to solid mechanics. The authors investigated linear elastic problems and extended the formulation for elasto-plastic behavior. See also \cite{Arora2021}.
\citet{Zhang2022} presented a framework based on PINNs for identifying unknown geometric and material parameters. The authors compared their predictions for materials with internal voids/inclusions for the case of linear elasticity, hyperelasticity, and plasticity. 
\citet{Abueidda2022} combined deep energy method formulation and DL to solve the governing PDEs of the deformation in hyperelastic and viscoelastic materials. The authors show the capability of the method to accurately capture the three-dimensional mechanical response without requiring any time-consuming training data. 

Many of the mentioned contributions are applied to homogeneous material systems. According to our investigation, they are not very suitable for heterogeneous systems, where we have a sharp contrast between different phases. Therefore, further refinements and extensions are required.
\citet{Rao2021} presented a PINN with mixed-variable output to model elastodynamics problems without resorting to labeled data, where the initial and boundary conditions are hardly imposed. In other words, the network design automatically satisfies the particular given BCs. The authors considered both the displacements and stress components as the network's outputs. Interestingly, this idea was also investigated before in the hybrid FE analysis. As a result, the accuracy and trainability of the network improved significantly.  
\citet{Yadav} solved a series of problems in solid mechanics using a domain-distributed version of distributed PINN and demonstrated that the approach is capable of addressing volumetric locking or capturing discontinuities. 
\citet{Henkes2022} addressed non-homogeneous materials which can trigger the nonlinearities in the stress and displacement fields. They have also utilized domain decomposition to get more accurate solutions. See also investigations by \cite{JAGTAP2020113028, ZHANG2021100220, Zhang2020PhysicsInformedNN}.
It is worth mentioning that the methodologies above can be categorized as mesh-free approaches in numerical modeling. Moreover, as we will explain through this work, collocation points at which the loss functions are minimized, can be interpreted similarly to nodes in finite element methods.

Similar to other numerical solvers, PINNs have the potential to be utilized for addressing multiphysics problems where we have a multi-objective optimization problem to satisfy several coupled governing equations.
\citet{He2020} employed PINNs in subsurface transport problems to minimize the governing equations of the problem in addition to loss functions based on the measurement data. \citet{Nguyen2022} utilized PINNs and the domain decomposition method to solve coupled problems for some benchmarks test, and they showed the capability of the model when it comes to a limited amount of data. \citet{PATEL2022} introduced a method for shock hydrodynamics that can satisfy hyperbolicity. \citet{GOSWAMI2020} developed a network for the phase-field model of the fracture (see also \citet{GOSWAMI2022114587}).
\citet{Amini2022} investigated the application of PINNs to the solution of problems involving thermo-hydro-mechanical processes.
\begin{figure}[H] 
  \centering
  \includegraphics[width=0.95\linewidth]{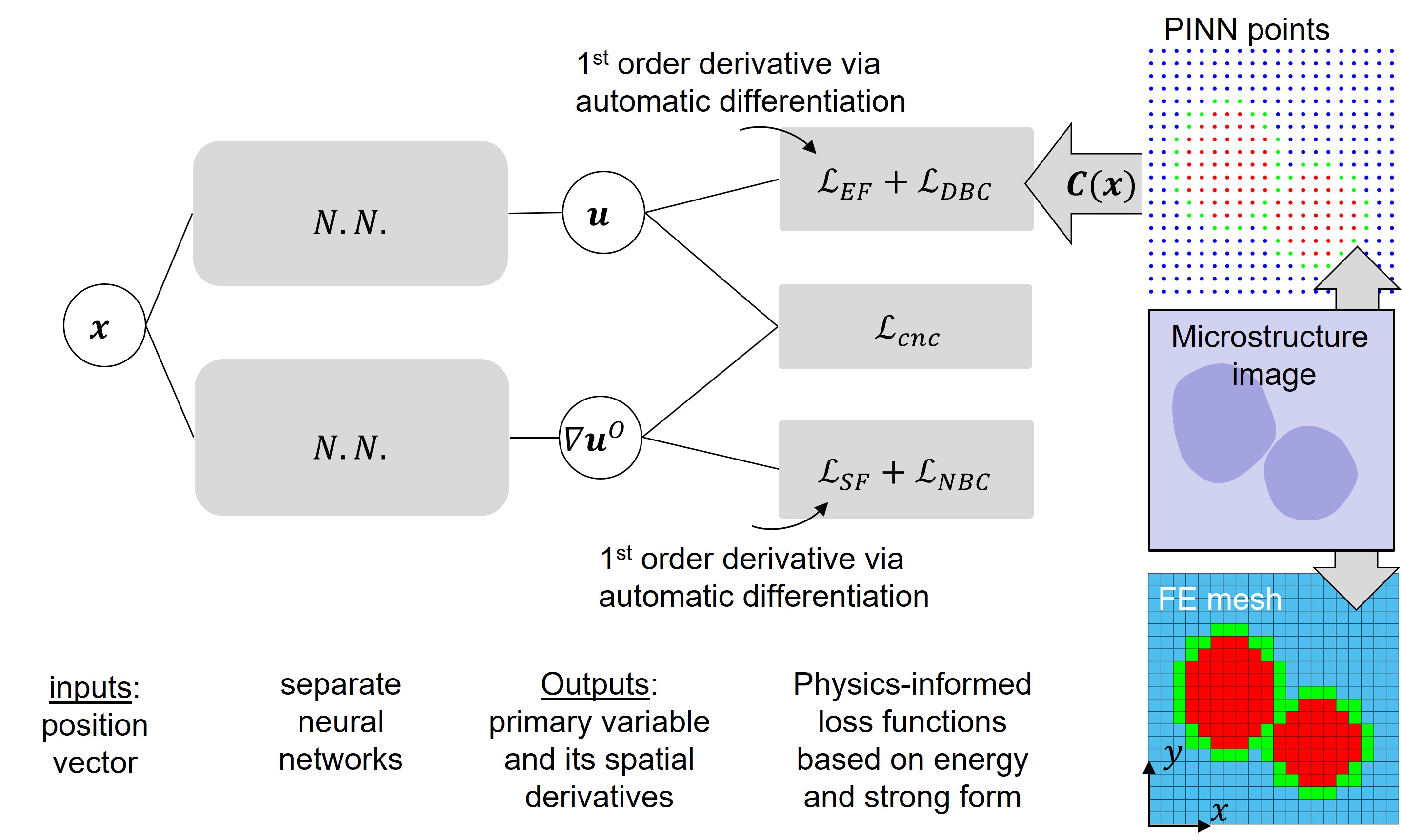}
  \caption{Proposed network architecture for solving a given boundary value problem. The vector $\bm{u}$ stands for any unknown primary variable (solution) that we are looking for. The idea is to combine the weak and strong forms of the problem simultaneously using separate networks for the primary variable and its spatial gradients. The networks are finally connected by means of an additional loss function.}
  \label{fig:intro}
\end{figure}

This work will discuss a simple yet effective extension of the PINN approach to directly solve a given PDE and boundary condition in a heterogeneous domain (see Fig.~\ref{fig:intro}). In section 2, the formulation of the problem for linear elasticity and steady-state diffusion problem is described. For the diffusion problem, we mainly focused on the heat transfer but it can be directly interpreted as an electrostatic, electric conduction, or magnetostatic problem. The derivation of the energy form from the PDE is also shortly reviewed for each of the described physics to show its applicability to other similar problems. In section 3, the architecture and details of the feed-forward neural network are described where we summarize all the necessary loss terms. This work is implemented using the SciANN package \citet{SciANN} but the methodology can easily be transferred to other platforms for similar studies. Finally, we provide a conclusion and future ideas on how this work can be extended for fast and accurate prediction of material behavior in multi-scaling scenarios.

\section{Formulation of the problem}
\subsection{Linear elasticity in heterogeneous solids}
In the context of linear elasticity, the formulation of the mechanical problem is described by three main equations: kinematics, constitutive relation (Hooke's law), and mechanical equilibrium. The kinematic relation defines the strain tensor $\bm{\varepsilon}$ in terms of the deformation vector $\bm{u}$ and reads
\begin{align}
\label{kinematics}
\bm{\varepsilon}\,=\,\text{sym}(\text{grad}(\bm{u})) = \dfrac{1}{2}\left(\nabla \bm{u} + \nabla \bm{u}^T \right).
\end{align} 
Employing the fourth order elasticity tensor $\mathbb{C}$, the second order stress tensor reads
\begin{align}
\label{materiallaw4}  
{\bm{\sigma}}\,={\mathbb{C}}(x,y)~{\bm{\varepsilon}}.
\end{align}
Note that we are, in particular, interested in a heterogeneous domain where material properties may vary throughout the microstructure. The latter explains the dependency of the stiffness tensor on the coordinates of collocation points. The mechanical equilibrium in the absence of body force as well as the Dirichlet and Neumann boundary conditions are written as 
\begin{align}
\label{Equilbrium}
\text{div}({\bm{\sigma}})\ &= \bm{0}~~~\text{in}~~~ \Omega\\
\label{BcsMech_d}
\bm{u} &= \bar{\bm{u}}~~~\text{on}~~\Gamma_D \\ 
\label{BcsMech_n}
\bm{\sigma} \cdot \bm{n} = \bm{t} &= \bar{\bm{t}}~~~~\text{on}~~\Gamma_N
\end{align} 
In the above relations, $\Omega$ and $\Gamma$ denote the material points in the body and on the boundary area, respectively.
For convenience, the second-order strain tensor in Eq.~\ref{materiallaw4} is usually written in the Voigt notation as
\begin{align}
\label{materiallaw}
\hat{\bm{\sigma}}\,=\hat{{C}}(x,y)~\hat{\bm{\varepsilon}}.
\end{align} 
In Eq.~\ref{materiallaw}, $\hat{\bullet}$ denotes the tensor ($\bullet$) in the Voigt notation. Considering the plane strain assumption in a two dimensional setting, we write the position-dependent elasticity tensor $\hat{\mathcal{C}}$ as (see also Fig.~\ref{fig:intro}):
\begin{align}
\label{elasticityPlanestrain}
\hat{{C}}(x,y)\,=\, \dfrac{E(x,y)}{(1-2\nu(x,y))(1+\nu(x,y))}\,\begin{bmatrix} 1-\nu(x,y) & \nu(x,y) & 0 \\
\nu(x,y) & 1-\nu(x,y) & 0 \\
0 & 0 & \dfrac{1-2\nu(x,y)}{2}
\end{bmatrix}.
\end{align}
Here, Young's Modulus $E$ and Poisson’s ratio $\nu$ represent the elastic constants for the material. We assume that the material behavior remains isotropic in each phase. Nevertheless, one can also focus on fully anisotropic material by defining $\hat{\mathcal{C}}$ differently and the formulation stays the same.
To obtain the Galerkin weak form (WF) of the problem we first multiply Eq.~\ref{Equilbrium} by a test function $\delta \bm{u}$, which satisfies the Dirichlet boundary conditions of the given boundary value problem. By applying integration by parts and utilizing Gauss theorem one obtains
\begin{align}
\label{Weakform}
\int_{\Omega} \delta{\bm{\hat{\varepsilon}}}^T\,\hat{{{C}}}(x,y)\,\hat{\bm{\varepsilon}}\,~dV\,-\, \int_{\Gamma_N} \delta{\bm{u}^T}\,\bar{\bm{t}}~dA=0. 
\end{align} 
The WF in Eq.~\ref{Weakform} leads to another representation which is based on the balance between the mechanical internal energy $E^M_{\text{int}}$ and mechanical external energy $E^M_{\text{ext}}$. In other words, we can also represent the problem based on the minimization of the total potential energy in which we have the contribution of internal and external forces (see also \cite{SAMANIEGO2020112790}).
\begin{align}
\label{int_Energy}
\text{Minimize:~}&\underbrace{\int_{\Omega}\dfrac{1}{2}\hat{\bm{\varepsilon}}^T\,\hat{{{C}}}(x,y)\,\hat{\bm{\varepsilon}}\,~dV}_{E^M_{\text{int}}} - 
\underbrace{\int_{\Gamma_N} {\bm{u}^T}\,\bar{\bm{t}}~dA}_{E^M_{\text{ext}}},\\
\label{int_Energy_bcs}
\text{Subjected to:~}& \text{BCs in Eq.~\ref{BcsMech_d} and Eq.~\ref{BcsMech_n}}.
\end{align}
The expressions in Eqs.~\ref{int_Energy} and \ref{int_Energy_bcs} will be added directly to the loss function of the neural network for the primary variable $\bm{u}$ (see also \cite{SAMANIEGO2020112790}).
%

\subsection{Diffusion Problem in heterogeneous solids}
We consider another prototype problem: the steady-state diffusion problem in a $2$-D setting. The equations are described for the problem of heat transfer. Note that the same derivation can be applied to study other similar fields such as electrostatic, electric conduction, or magnetostatic problem \cite{Guo2022}. 
The heat flux $\bm{q}$ is defined according to Fourier's law:
\begin{align}
\label{Fourier}
\bm{q} = -k(x,y)\,\nabla T. 
\end{align} 
Here $k(x,y)$ is the position-dependent heat conductivity coefficient. The governing equation for this problem is based on the heat balance in the absence of heat source and reads
\begin{align}
\label{StrongfromThermal}
\text{div}(\bm{q}) &= 0~~~~~~ \text{in}~ \Omega, \\
\label{BCsthermalD}
T &= \bar{T}~~~~~\text{on}~ \Gamma_D, \\
\label{BCsthermalN}
\bm{q}\cdot \bm{n} = q_n &= \bar{q}~~~~~~\text{on}~ \Gamma_N.
\end{align} 
In the above relations, the Dirichlet and Neumann boundary conditions are introduced in Eq.~\ref{BCsthermalD} and Eq.~\ref{BCsthermalN}, respectively. Similar to the procedure in the mechanical problem, by introducing $\delta T$ as a test function, the weak form of the steady-state diffusion problem reads
\begin{align}
\label{weakformthermal}
\int_{\Omega}\,k(x,y)\,\nabla^T T\,\delta(\nabla T)~dV\,+\,\int_{\Gamma_N}\bar{q}~\delta T~dA\,=\,0.
\end{align}
Analogously to the previous section, the weak form of the thermal problem can be interpreted as the variation of the energy and allows us to introduce an equivalent balance between the thermal internal energy $E^T_{\text{int}}$ and the thermal external energy $E^T_{\text{ext}}$. Therefore, we reformulate the problem as an energy minimization: 
\begin{align}
\label{int_Energy_ther}
\text{Minimize:~}&\underbrace{\int_{\Omega}\dfrac{1}{2}\,k(x,y)\,\nabla^T T\,\nabla T\,~dV}_{E^T_{\text{int}}} + 
\underbrace{\int_{\Gamma_N} \bar{q}\,T~dA}_{E^T_{\text{ext}}},\\
\text{Subjected to:~}& \text{BCs in Eq.~\ref{BCsthermalD} and Eq.~\ref{BCsthermalN}}.
\end{align}
In the expression for the loss function, the above term is evaluated for the primary variable $T$. 

\begin{figure}[H] 
  \centering
  \includegraphics[width=0.75\linewidth]{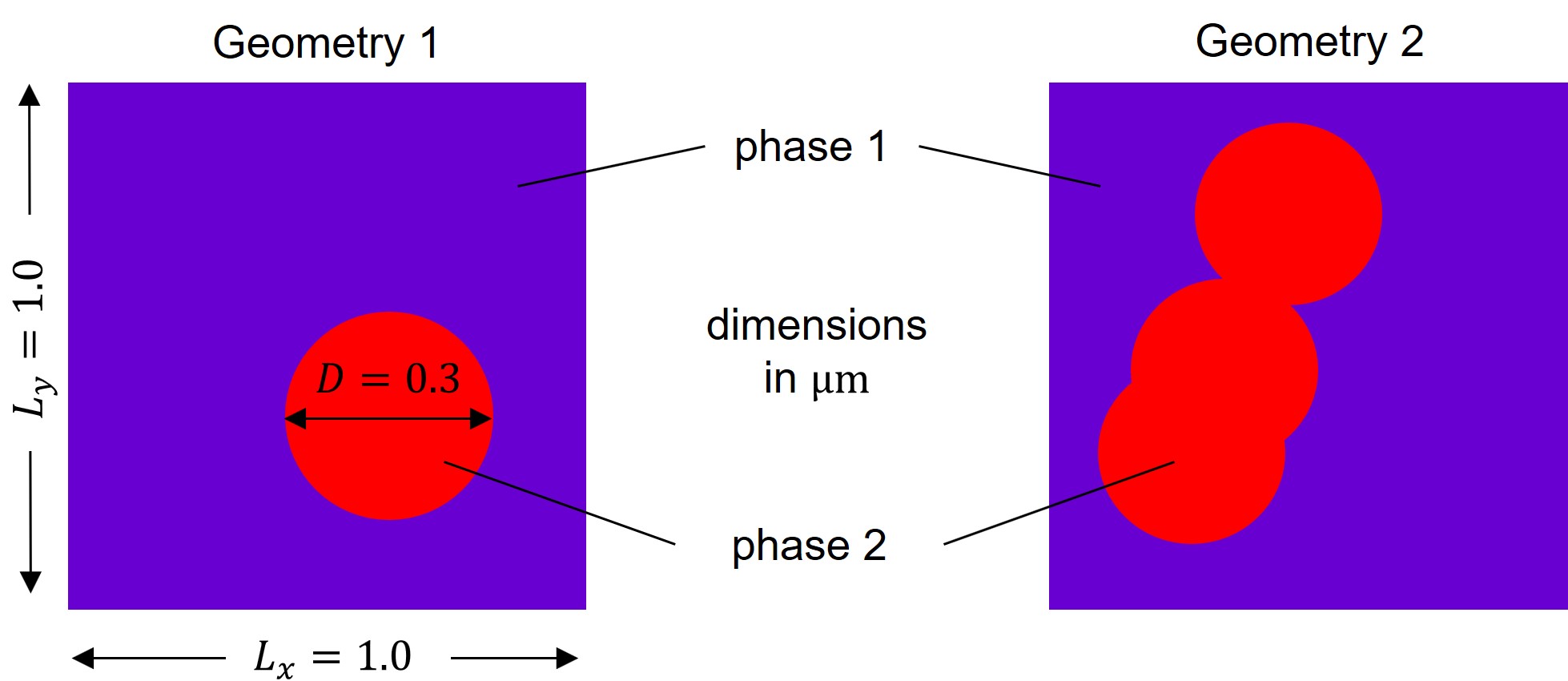}
  \caption{Selected geometries for studies on mechanical and steady-state diffusion problem.  }
  \label{fig:allgeom}
\end{figure}

\begin{figure}[H] 
  \centering
  \includegraphics[width=0.8\linewidth]{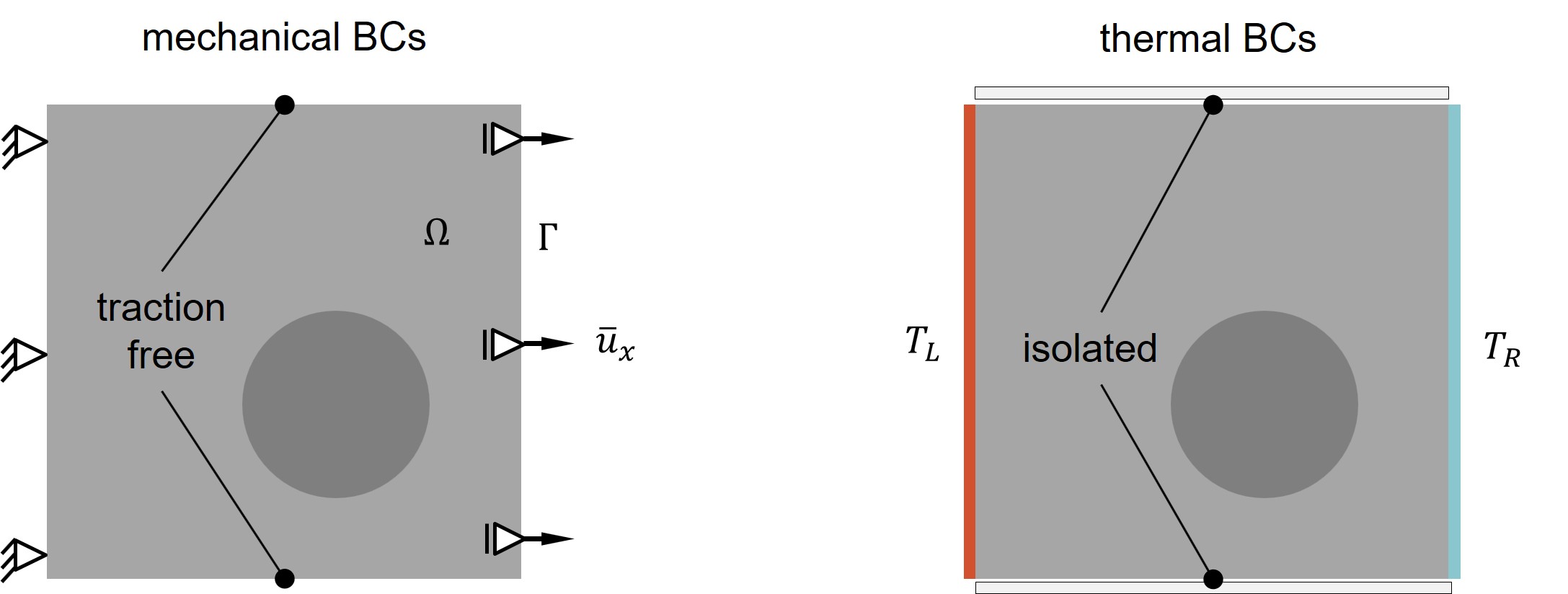}
  \caption{Boundary conditions for mechanical and thermal problems.}
  \label{fig:BC}
\end{figure}

In the current study, we study two different set of geometries according to Fig.~\ref{fig:allgeom}. They are selected to cover some important cases which happen in different engineering problems. In addition to phase changes (geometry 1), sharp corners between different phases are introduced in geometry 2. Material properties are varied for different phases. The mechanical and thermal problem described in section~\ref{sec:mech} and section~\ref{sec:ther}, respectively. For these two problems, we change the Young's modulus and thermal conductivity of the involved material according the relation $E_{\text{inc}}/E_{\text{mat}} = k_{\text{inc}}/k_{\text{mat}} = 2.0$. Later on, The phase contrast as well as BCs are also varied to investigate the PINN performance (see section \ref{sec:mix}).
For the mechanical problem, we apply a tensile test while the upper and lower side of the specimen are traction free. For the thermal problem, we apply constant temperature at the left and right edge as shown in the right-hand side of Fig.~\ref{fig:BC}. The upper and lower part are isolated. The material parameters as well as the model inputs are summarized in Table~\ref{tab:par}.
\begin{table}[H]
	\centering
	\begin{tabular}{ l l  } \hline
	\multirow{1}{*}{}         & Value/Unit    \\ \hline \hline 
	\textbf{Mechanical problem (static equilibrium)} \\
	Phase 1 Young's modulus and Poisson's ratio ($E_{\text{mat}}$, $\nu_{\text{mat}}$)  & ($0.5$ GPa, $0.3$)  \\
	Phase 2 Young's modulus and Poisson's ratio ($E_{\text{inc}}$, $\nu_{\text{inc}}$)  & ($1.0$ GPa, $0.3$)  \\
	Applied displacement in $x$-direction, $\bar{u}_x$ & $0.05$~$\mu$m\\
	\hline
	\textbf{Diffision problem (heat conduction)} \\
 	Phase 1 heat conductivity ($k_{\text{mat}}$)  & $1.0$~W/mK  \\
 	Phase 2 heat conductivity ($k_{\text{inc}}$)  & $0.5$~W/mK  \\
	Applied temperature ($T_L$,$T_R$)  & ($1.0$~K, $0.0$~K)\\
    \hline
	\end{tabular}
	\caption{Model input parameters for mechanical and thermal problems.}
	\label{tab:par}
\end{table}

\newpage
\section{New architecture for PINNs}


\subsection{Mechanical problem} \label{sec:mech}
In the PINN, the output variables (i.e. displacement and/or stress) are approximated via fully connected feed-forward neural networks (FFNN) \cite{RAISSI2019}. The network has three types of layers. The very first layer is the input layer, next we have several hidden layers, and finally comes the output layer \cite{schmidhuber2015deep}. Each layer can contain several neurons, and there is no connection between neurons inside a layer. Moreover, each layer is connected to the next layer to pass the information from the input layer to the output layer. For example, let's denote information within the $l$-th layer by the vector $\bm{z}^l$ where each component of the vector represents the value stored in that particular neuron. Therefore, transferring the information between two neighboring layers within a FFNN can be written as:
\begin{equation}
\label{eq:NN_1}
    z^l_m = a (\sum_{n=1}^{N_l} w^l_{mn} z_n^{l-1} + b^l_{m} ),~~~l=1,\ldots,L. 
\end{equation}
Here the network has $L$ layers and $N_l$ neurons in the $l$-th layer. The connection between the \emph{n}-th neuron in layer $l-1$ and \emph{m}-th neuron in layer \emph{l} has a weight $w_{mn}$. For each neuron in the $l~$th hidden layer, we have a bias term $b^l_m$. The input layer in this work is the components of the position vector in 2D, i.e. we have $z^0_1=x$, and $z^0_2=y$. Moreover, the output layer contains the desried solution and therefore is written as $z^L_1=u_x$, $z^L_2=u_y$, $z^L_3=\sigma_x$, $z^L_4=\sigma_y$ and $z^L_5=\sigma_{xy}$.
Each neuron presents a summation operation (so-called weighted sum) plus a bias term and passes it through a non-linear activation function $a(\cdot)$. The activation function utilized in this work is Hyperbolic-Tangent (tanh) function \cite{sibi2013analysis}: 
\begin{equation}
    \tanh(x)=\frac{\exp^x - \exp^{-x}}{\exp^x + \exp^{-x}}.
\end{equation}
The network trainable and input parameters are also shown by $\bm{\theta}=\{\bm{W},\bm{b}\}$ and $\bm{X}=\{x,y\}$, respectively. Here, $\bm{W}$ and $\bm{b}$ consist of all the weights and biasis in neural network (their components for the $l~$the layer are also shown in Eq.~\ref{eq:NN_1}). Finally, by denoting each neural network system by $\mathcal{N}$, we write the following solution components for the problem:
\begin{align}
\label{eq:NN_3}
    u_x           &= \mathcal{N}_{u_x} (\bm{X}; \bm{\theta}), \\
    u_y           &= \mathcal{N}_{u_y} (\bm{X}; \bm{\theta}), \\
    \sigma^O_x    &= \mathcal{N}_{\sigma_x} (\bm{X}; \bm{\theta}), \\
    \sigma^O_y    &= \mathcal{N}_{\sigma_y} (\bm{X}; \bm{\theta}), \\
    \sigma^O_{xy} &= \mathcal{N}_{\sigma_{xy}} (\bm{X}; \bm{\theta}).
\end{align}
To build a given partial differential equation residual and its relative boundary conditions, one requires the differentiation of the output layer (e.g. displacement and stress) with respect to the input layer (e.g. position vector). In deep neural networks, the analytical differentiation is already available by using the automatic differentiation (AD) \cite{baydin2018automatic}, which is a feature in modern deep learning frameworks such as Pytorch \cite{paszke2017automatic} and Tensorflow \cite{abadi2016tensorflow}.

Based on the described set of equations for the mechanical problem (Eqs.\ref{kinematics} to \ref{BcsMech_n} as well as  Eqs.~\ref{int_Energy} and \ref{int_Energy_bcs}), the following loss functions are defined below. As discussed earlier, the new architecture consists of three main parts. First, we have the loss terms applied to the displacement vector $\bm{u}$. These terms are based on the energy form of the problem as well as the Dirichlet BCs ($\mathcal{L}_{EF}$ and $\mathcal{L}_{DBC}$). Next, we add the loss terms applied to the stress tensor $\bm{\sigma}^o$ as well as the Neuman BCs ($\mathcal{L}_{SF}$ and $\mathcal{L}_{NBC}$). Finally, we have the connection loss term $\mathcal{L}_{cnc}$.  
\begin{align}
\label{Totalloss}
\mathcal{L} &= \underbrace{\mathcal{L}_{EF} + \mathcal{L}_{DBC}}_{\text{based on $\bm{u}$}} + \mathcal{L}_{cnc} + \underbrace{\mathcal{L}_{SF}  + \mathcal{L}_{NBC}}_{\text{based on $\bm{\sigma}^o$}}, \\
\label{loss_weak}
\mathcal{L}_{EF} &= \text{MAE}_{EF}\left( -\int_{\Omega}^{} \frac {1}{2} {\bm{\sigma}} : {\bm{\varepsilon}} \,dV + \int_{\Gamma}^{} ({\bm{\sigma}} \cdot \bm{n})\bm{u} \,dA \right), \\
\label{loss_DBC}
\mathcal{L}_{DBC} &= \text{MSE}_{DBC}\left( \bm{u} - \overline{\bm{u}} \right), \\
\label{loss_cnc}
\mathcal{L}_{cnc} &= \text{MSE}_{cnc}\left( \bm{\sigma}^o - \bm{\sigma} \right), \\
\label{loss_SF}
\mathcal{L}_{SF} &= \text{MSE}_{SF}\left( \text{div}(\bm{\sigma}^o) \right), \\
\label{loss_NBCx}  
\mathcal{L}_{NBC} &= \text{MSE}_{NBC}\left(\bm{\sigma}^o \cdot \bm{n} - \overline{\bm{t}} \right).
\end{align}
The energy loss (Eq.~\ref{loss_weak}) is supposed to be minimized and it takes at the global response of the system, whereas the other loss terms are satisfied locally for each collocation point. 
Furthermore, we have the definition of the mean squared error according to
\begin{align}
\label{eq:MSE}
\text{MSE}( \bullet )_{type} &= \dfrac{1}{k_{type}}\sum_{i=1}^{k_{type}}(\bullet)^2,~~~
\text{MAE}( \bullet )_{type} = \dfrac{1}{k_{type}}\sum_{i=1}^{k_{type}}|\bullet|,\\
k_{EF}=1,~~~k_{DBC}&=N_{DBC},~~~k_{cnc}=N_{b},~~~k_{SF}=N_{b},~~~k_{NBC}=N_{NBC}.
\label{eq:type}
\end{align}
Here, $k_{type}$ is the number of collocation points for each loss term. Note that, we have different number of collocation points for each boundary type (i.e. $N_{DBC}$ and $N_{NBC}$) and bulk ($N_{b}$). The above relations are further exploited in Appendix C for described BVPs in Fig.~\ref{fig:BC}. 

The solution to the mechanical problem is calculated by minimizing the total loss function in Eq.~\ref{Totalloss} on the set of collocation points. The final optimization reads as:
\begin{align}
\label{minimize}
\bm{\theta}^* = \arg \min_{\bm{\theta}} \mathcal{L}(\bm{X}; \bm{\theta}).
\end{align}
In the above equation, $\bm{\theta}^*$ includes the optimal trainable network parameters. The collocation points $\bm{X}$ are located inside the domain and on the boundary.

In this work, the Adam optimizer \cite{kingma2014adam} is employed. Several parameters and hyper-parameters affect the network’s performance, such as activation function, learning rate, number of epochs, batch size, number of hidden layers, and number of neurons in each layer. Their influence should be investigated based on the number of collocation points and the network’s architecture. This matter is explained in more detail in Section \ref{sec:hype}.\\

\textbf{Remark 1} One may consider different weightings for the terms in Eq.~\ref{Totalloss}. Our preliminary studies show that in some cases, by tuning the weightings, one can improve the results for the same number of iterations in training. In the current work, we considered equal weightings.\\


\textbf{Remark 2} In Eq.~\ref{loss_weak} we considered mean absolute error (MAE) while for the other terms, mean square error (MSE) is utilized. Note that in the energy formulation, the magnitude of the loss term is much smaller compared to other terms (see the multiplication of stress and strain tensor). According to our investigation and as it is discussed in Section \ref{sec:1st_example}, using MAE for the energy term is more beneficial to keeping the values of different loss functions in the same order.  \\

\begin{figure}[H] 
  \centering
  \includegraphics[width=1.0\linewidth]{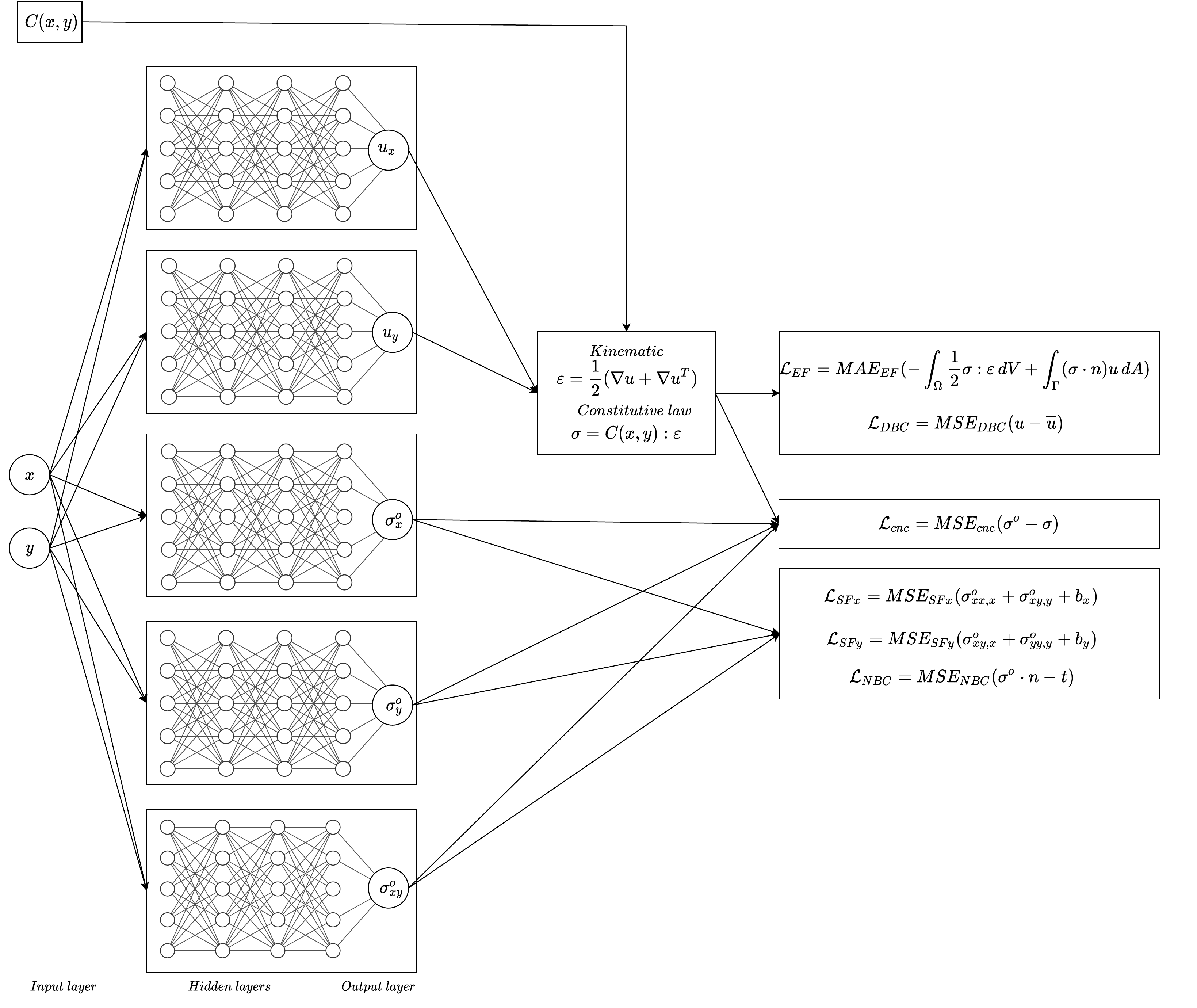}
  \caption{ Architecture and loss functions for the mixed formulation based network in the mechanical problem.  }
  \label{fig:PINNs_net_m}
\end{figure}

The architecture of the network for the mechanical problem is shown in Fig.~\ref{fig:PINNs_net_m}. The network is based on the Eqs.~\ref{Totalloss} to \ref{loss_NBCx} which shows up on the right-hand side of the figure. The proposed network in this study shares some similarities with the recent and previous investigation in the literature \cite{Henkes2022, SAMANIEGO2020112790, HAGHIGHAT2021, RAISSI2019}. Nevertheless, it also differs from them in the following aspects. First, the network takes components of the position vector as inputs ($x$ and $y$ coordinate in a 2D setting) and predicts the components of the deformation vector $\bm{u}$ and stress tensor $\bm{\sigma}^O$ as outputs. 
Using the predicted deformation vector components ($u_x$ and $u_y$) and based on the kinematic relation (Eq.~\ref{kinematics}) and constitutive law (Eq.~\ref{materiallaw4}), strain and stress tensors are calculated from automatic differentiation, respectively. The latter quantities are denoted by $\bm{\varepsilon}$ and $\bm{\sigma}$. This stress tensor is evaluated from deformation $\bm{u}$ and it is in principle different from the stress tensor $\bm{\sigma}^O$ which is the direct output of the network. Later on, these quantities are forced to take the same value utilizing the connection loss term $\mathcal{L}_{cnc}$ presented in Eq.~\ref{loss_cnc}. Next, in the new design, we consider separate networks for each of the output variables. According to our investigations which will be reported later on in this work, having separated networks for each output quantity shows some advantages compared to the case where all the networks are combined. 
Note that for constructing the physical constraints (i.e. loss function) we only need first-order derivatives of the output variables with respect to the input parameters. Reducing the order of derivatives also seems to be beneficial in different aspects. First, one has more flexibility in choosing the activation functions (i.e. we avoid vanishing its gradients by taking higher-order derivatives). Second, the computational cost is decreased compared to the cases where we have second-order derivation. Moreover, as we will show later on, it seems that lower-order derivatives result in more accurate predictions (less overall error). In simple studies in Appendix B, we indicate that the lower the degree of derivation is, the easier it is for the network to find the solution for a given number of iterations and collocation points.

The proposed network’s design is achieved by performing and comparing several computations for different network architectures. The summary of considered models (network architectures) is depicted in Table~\ref{tab:model}. The current work is mainly based on model E in which multiple outputs for displacement and stress quantities exist. Note that in models B, C and E only first-order derivatives of the outputs are required for obtaining the loss functions. 
\begin{table}[H]
\centering
\caption{Summary of possible models (network architectures) which can be used for PINNs. The inputs are always the same (position), while outputs and loss functions (LFs) are changed concerning the architecture. The current work is based on model E.} \label{tab:model}
\begin{footnotesize}
\begin{tabular}{l l }
\hline
Name &  Description    \\
\hline
Model A    &  outputs: components of $\bm{u}$, LFs: governing PDE and BCs, separated NNs are used \\ 
Model B    &  outputs: components of $\bm{u}$, LFs: energy form of the PDE and BCs, separated NNs are used \\ 
Model C    &  outputs: components of $\bm{u}$ and $\bm{\sigma}$, LFs: governing PDE and its energy form, combined NN is used \\
Model D    &  outputs: components of $\bm{u}$ and $\bm{\sigma}$, LFs: governing PDE, separated NNs are used \\ 
Model E    &  outputs: components of $\bm{u}$ and $\bm{\sigma}$, LFs: governing PDE and its energy form, separated NNs are used \\ 
\hline
\end{tabular}
\end{footnotesize}
\end{table}

The network parameters are summarized in Table~\ref{tab:network}. For other models, similar parameters are utilized except for the architecture of the inputs and outputs which is according to Table~\ref{tab:model}. In addition to the learning rate parameter $\alpha$, we have other parameters for the better stability and convergence of the Adam optimizer. The first moment is set to $0.9$ while the second moment of the gradient is set to $0.999$ (see also \cite{kingma2014adam}).
\begin{table}[H]
\centering
\caption{Summary of the network parameters.}  
\label{tab:network}
\begin{footnotesize}
\begin{tabular}{ l l }
\hline
Parameter                          &  Value    \\
\hline
Inputs, Outputs                  &  $\bm{X}$, ($\bm{u}$, $\bm{\sigma}^O$) \\ 
activation function                &  $\tanh$ \\ 
Number of layers and neurons per layer ($L$, $N_l$)  &  (5, 40) \\
Batch size                         &  full batch \\ 
(Learning rate $\alpha$, number of epochs)  &  $(10^{-3},10^{5})$ \\ 
\hline
\end{tabular}
\end{footnotesize}
\end{table}

\subsubsection{Single heterogeneity in a matrix (geometry 1)} \label{sec:1st_example}
\begin{figure}[H] 
  \centering
  \includegraphics[width=1.0\linewidth]{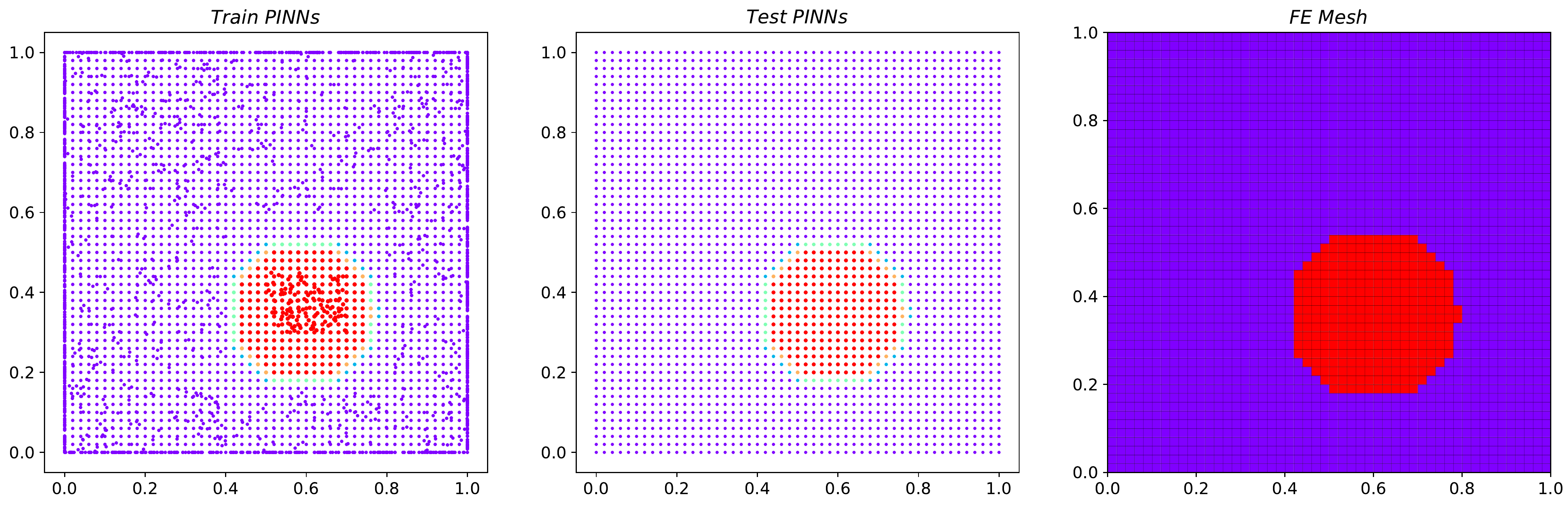}
  \caption{Left: collocation points utilized in the introduced PINN solver. Middle: points at which the trained network is evaluated. Right: quadrilateral meshes to discretize and solve the problem with finite element method. Different colors represent changes in the Young's modulus of the material.}
  \label{fig:geom1_config}   
\end{figure}

\begin{figure}[H] 
  \centering
  \includegraphics[width=0.95\linewidth]{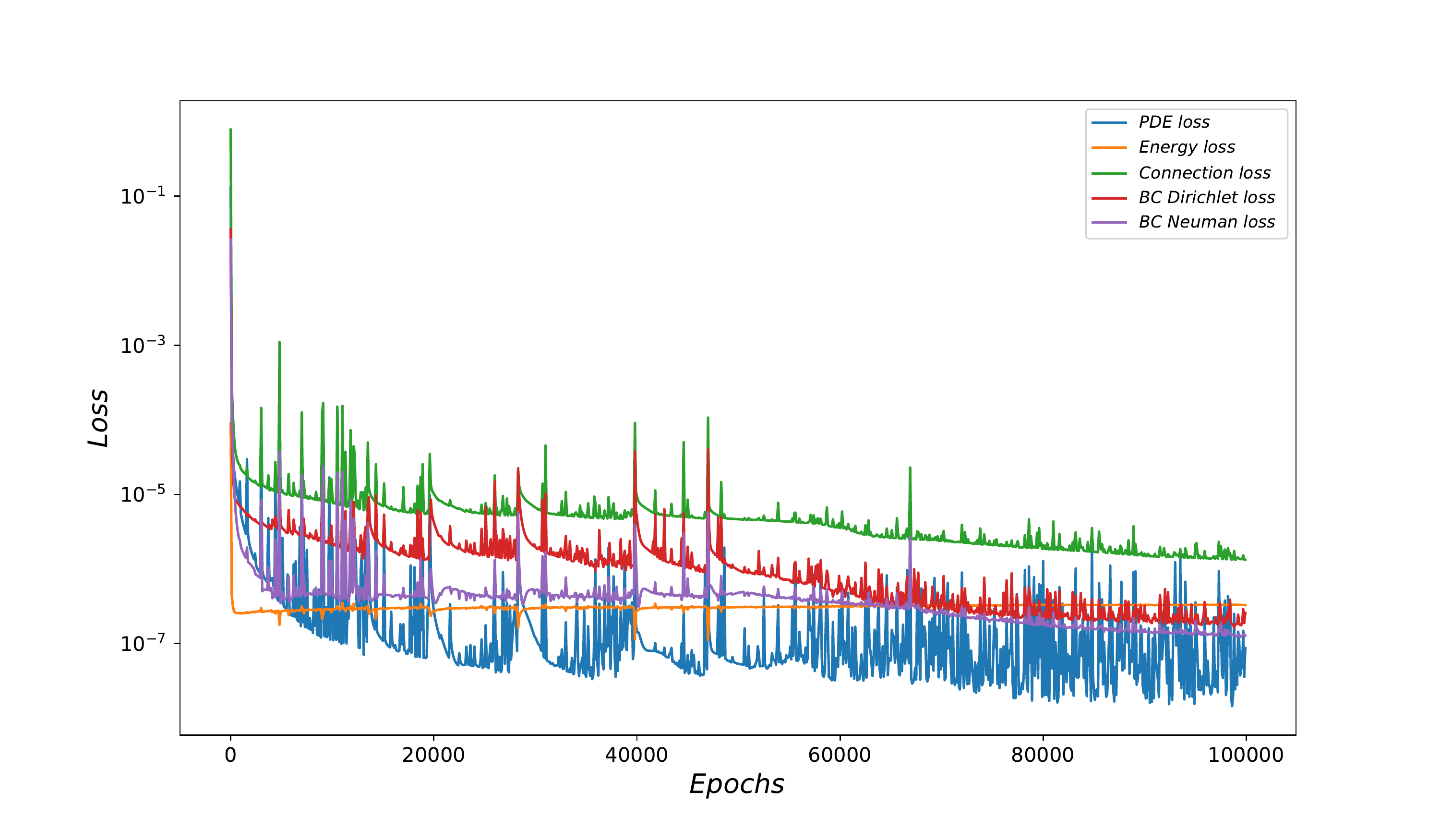}
  \caption{Individual loss terms versus number of epoch for geometry 1 using model E. }
  \label{fig:all_losses}
\end{figure}

To train the network (i.e. minimizing the introduced loss functions), collocation points are defined according to the left-hand side of Fig.~\ref{fig:geom1_config}. As we will discuss later, to improve the results, we consider more points (in a random way) within the red phase and the purple phase. Later on we evaluate the trained network for the test points shown in the middle of Fig.~\ref{fig:geom1_config}. The total number of collocation points is around $5000$. On each boundary, we considered $400$ points.  
Note that for a better comparison between FE and PINN, the FE meshes are defined regularly. Moreover, the same number of elements is considered for FEM (on the right-hand side) as points for PINN (in the middle of the figure). Colors indicate Young's modulus value (see Table~\ref{tab:par}). The Young's modulus of the points at the interface is computed concerning neighboring elements materials in FE mesh. This strategy is utilized to mimic the assembly procedure of FEM and assigns the relevant stiffness to every point.
Different loss terms introduced earlier are now minimized with respect to free parameters of the network (i.e. weighting and biases). In Fig.~\ref{fig:all_losses}, all the loss terms are plotted versus the number of epochs (iterations for the training).

\begin{figure}[H] 
  \centering
  \includegraphics[width=0.85\linewidth]{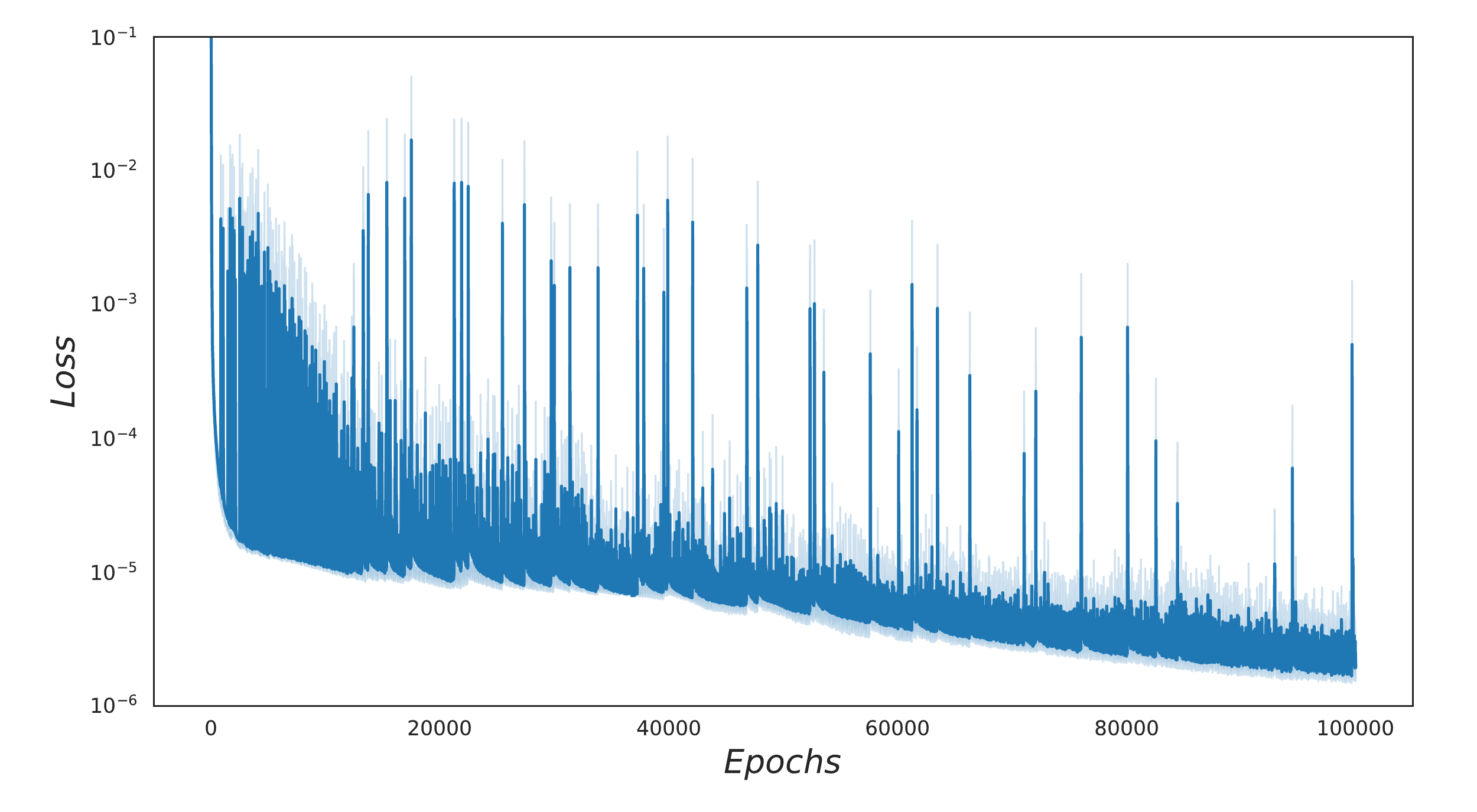}
  \caption{The total loss term together with the confidence bound 95 \% for geometry 1 using model E.}
  \label{fig:conf_bond}
\end{figure}

We minimized the same problem (same loss functions) for five different times to make sure that the introduced network is always able to minimize the total loss function. The main loss terms and how they decay through the training are plotted individually in Fig.~\ref{fig:all_losses}. We observe a very small values for the loss regarding the energy term that corresponds to MSE formulation. As we checked, by using absolute value error, the loss value will be in order of other loss terms but the final results did not changed significantly. 
The total loss value and its performance with $95 \%$ confidence bound are plotted in Fig.~\ref{fig:conf_bond}. 

We used Skylake CPU, Platinum 8160 model, and Intel HNS2600BPB Hardware Node Type. The computation for each BVP on averaged took around 2 hours, which is longer than the computation time from a FEM in the comparable hardware. This time can be reduced by using the GPU power, and it is not in our current scope to further elaborate on this point.

\begin{figure}[H] 
  \centering
  \includegraphics[width=1.0\linewidth]{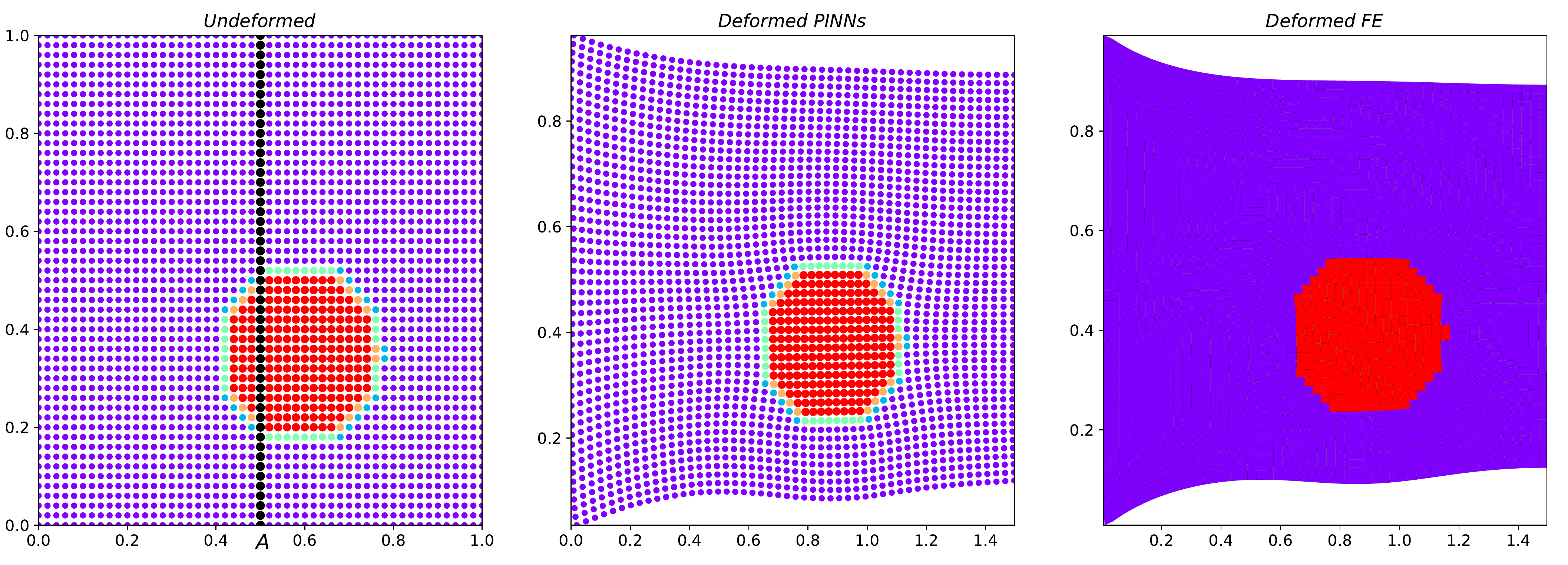}
  \caption{Undeformed and deformed configurations obtained from the mixed formulation based PINNs and FEM for the mechanical problem considering geometry 1. The results are magnified by a factor of $10$. }
  \label{fig:deformed_geom1}
\end{figure}

\begin{figure}[H] 
  \centering
  \includegraphics[width=0.95\linewidth]{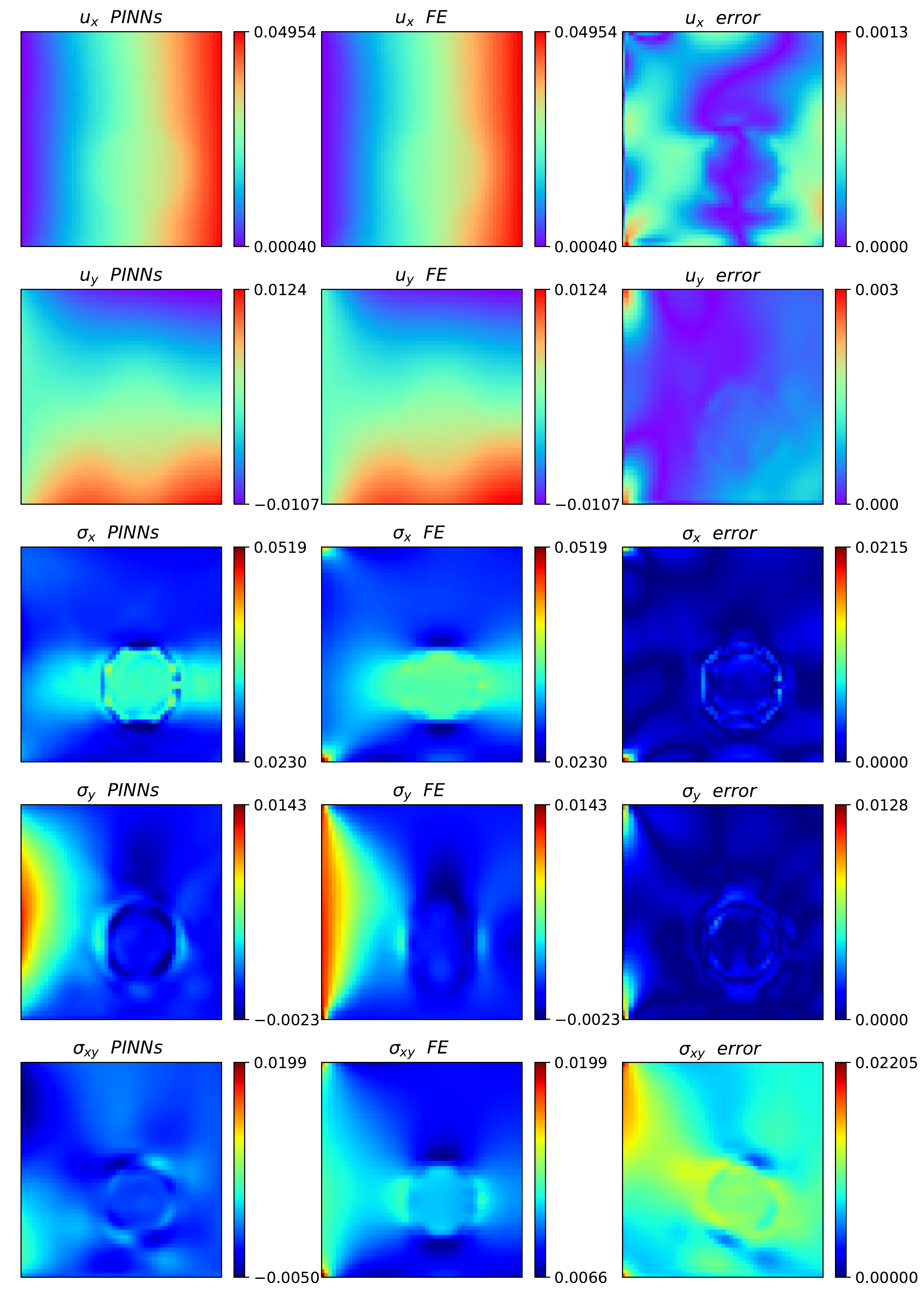}
  \caption{Obtained results for the mechanical problem using geometry 1 and the mixed formulation based PINNs and FEM. }
  \label{fig:res_geom1}
\end{figure}

The deformed configurations from the PINNs and the FEM are presented in Fig.~\ref{fig:deformed_geom1} where the deformations are magnified by a factor of $10$. The result match very well despite the small violation of boundary conditions on the left edge. Further results are reported in Fig.~\ref{fig:res_geom1} wherein the first two rows are the predicted displacement vector components and in the last three, the stress tensor components are presented. In the first, second, and third columns, the results from PINNs, FE, and the difference between these two methodologies are presented, respectively. 

In overall, for all the output values, we observe a good agreement between predictions of the network and the FE method. The difference between the results is bold first near to the corner points at the left boundary, and second, in the area where we have the sharp phase transition. The maximum relative difference which is localized in the critical spots are about 44 \% and 77 \% for displacement and stress, respectively. The averaged relative difference value is about 4 \% and 3.7 \% for displacement and stress values, respectively. Despite the acceptable performance for the normal components of the stress tensor, the prediction of the mixed formulation based PINNs for the shear component ($\sigma_{xy}$) is relatively poor. As we discuss in Section \ref{sec:ther} and Appendix B, the difference can be reduced by some simple remedies such as running the PINNs model for more epochs and adding more initial collocation points in critical spots such as the external boundary of the problem and interphase regions.


One may notice that for this example, the chosen points for evaluating the results and comparing them against those from FEM are shown in the middle part of Fig.~\ref{fig:geom1_config}. These points are mainly included in the collocation points (left side of Fig.~\ref{fig:geom1_config}). In the next study, we will examine this matter further for higher resolution (more test points and meshes) to make sure the obtained solution can be properly interpolated in the whole domain.

\begin{figure}[H] 
  \centering
  \includegraphics[width=1.0\linewidth]{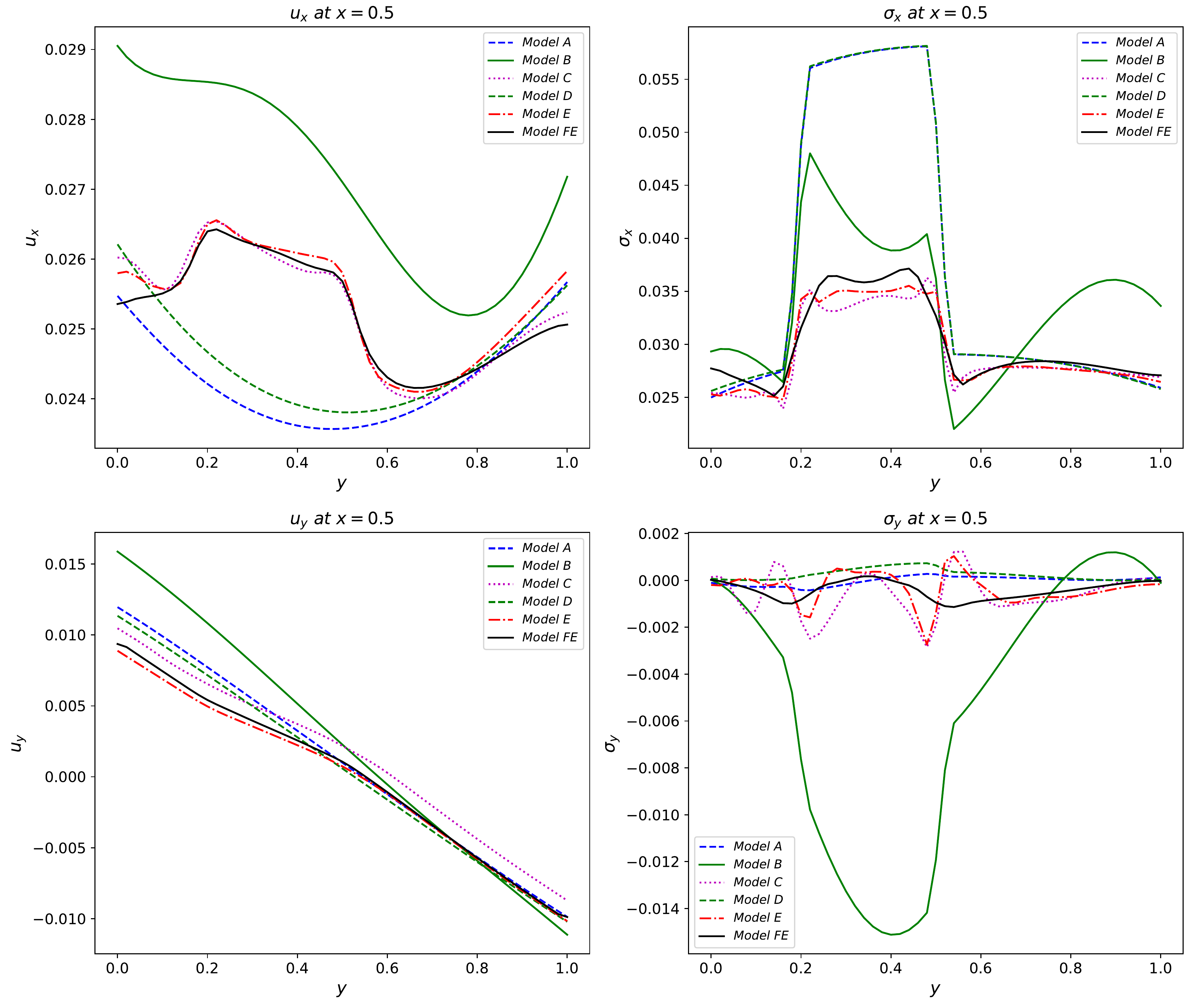}
  \caption{Comparison between obtained results from examined networks and FE along $x=0.5~\mu$m direction. For model Model A, outputs are components of $\bm{u}$ and loss functions are based on governing PDE as well as BCs. For Model B, outputs are components of $\bm{u}$ and loss functions are based on energy form of the PDE as well as BCs. For Model C, outputs are components of $\bm{u}$ and $\bm{\sigma}$ and loss functions are based on governing PDE and its energy form in a combined NN. For Model D, outputs are components of $\bm{u}$ and $\bm{\sigma}$ and loss functions are based on governing PDE where separated NNs are used. For Model E, outputs are components of $\bm{u}$ and $\bm{\sigma}$ and loss functions are based on governing PDE as well as its energy form where separated NNs are used. }
  \label{fig:comp_net_disp}
\end{figure}

According to Table~\ref{tab:model}, one has different options for designing the network architecture. In models A and D, the PDE is directly implied in the loss function. In these cases, we expect second-order derivation. The latter point can be a drawback as the network has to approximate the solution by taking higher order of derivatives compared to the previous case. In Appendix B, we show in a simple example that the lower the degree of derivation is, the easier it is for the network to find the solution. Therefore, utilizing the energy form of the problem can be beneficial (see also \cite{SAMANIEGO2020112790}). In models A and B, the only output of the network is the displacements $\bm{u}$ whereas in the rest of the models, components of the stress tensor are also considered as additional outputs. We would like to study this point and argue that having an additional set of outputs helps for better training in heterogeneous systems (see also \cite{HAGHIGHAT2021, Zhang2022, SAMANIEGO2020112790}). 
In model E (current work), in addition to separate outputs for displacements and stresses, also a separate network is proposed for the each output \cite{HAGHIGHAT2021}. To show the performance of each model, we utilized each of them on geometry $1$ under the same boundary conditions. For each training, $100$K epoch is used. The results for predicted $u_x$, $u_y$, $\sigma_x$, and $\sigma_{y}$ along the cross-section at $x=0.5~\mu$m are shown in Fig.~\ref{fig:comp_net_disp}.

Considering the results from the FE method as the reference solution, only models C and E were able to capture the response of the heterogeneous system properly. The other models were able to capture the overall behavior but many details are missed. The results of model B compared to A indicates that only reducing the derivation order is beneficial (see also Appendix B). Note that, in model B, we only minimize the total energy globaly, while in model A, the governing equations are satisfied pointwise. The latter argument can be an explanation for errors in model B. Comparing models A and D indicates that by only increasing the number of outputs, one may not achieve a significant improvement. On the other hand, by combing all these refinements together, more accurate response is achieved. The performance of model C seems to be less accurate compared to model E which indicates the advantage of using a separated network for each of the output variables. Note that utilizing separate NN for each output means that we used much more (roughly five times more) free variables for the network’s training. In the future, more studies are required to investigate the optimum design of the network architecture.

Based on the results in Fig.~\ref{fig:comp_net_disp}, we conclude that on average, the difference between the network predictions and the FE solution is higher for the stress values. Nevertheless, model E managed to capture the jump in the response properly. We believe that part of the difference is also because of the way we meshed the system for FE. In the future comparison, one can also employ a better discretization for the interface of the different phases. In the current work, we intend to stay with the so-called pixel-based mesh which might introduce unnecessary jumps in the prediction of the FE results. See the sharp corners in Fig.~\ref{fig:geom1_config} for the elements at the border of heterogeneity.

\textbf{Remark 3} The above comparison does not imply that models A, B, and D are not useful for finding the solution in a heterogeneous domain. One can do further studies and tune the number of neurons, layers, the number of epochs, and collocation points accordingly to get a better response out of each model. This also holds for model C. In general, having a fair comparison between all these models is not straightforward. The least we can conclude is that adding separate networks combined with the energy form can be beneficial for finding the solution in heterogeneous domains. For a better understanding readers are also encouraged to see Appendix B, where we performed studies on a simplified ordinary differential equations.

\subsubsection{Discussions on the parameters and hyper-parameters for deep learning model}\label{sec:hype}
In feed-forward neural networks, there are several training parameters and hyper-parameters which affect the optimization process and also the final value of the loss function. Therefore, to get the best performance from the neural network, one requires to perform a series of parameter studies during the training process. In the following we summarize our experience with choice of some hyperparameters.

\begin{itemize}
\item Activation functions have a significant influence on the network performance. Depending on the order of the PDE, the higher-order derivative of the active functions should be non-zero for the gradient descent algorithm to perform well. In this work, we tried various qualified activation functions such as \emph{tanh}, \emph{sigmoid}, and \emph{softplus}. From available options, \emph{tanh} showed the best performance in terms of the final value of the loss function. 

\item It is crucial to use sufficient epochs. In the process of training, when the network sees all of the collocation points, then the network is trained for one epoch. One can also define a criterion to stop the training based on the total loss. However, in this work, we set the stopping value as $10^{-12}$, and we train our networks for $10^5$ epochs.

\item We can feed all the collocation points for the network training at the same time in one step, or we can divide them into several batches or mini-baches. By choosing mini-baches, we have the option to shuffle the data in each epoch. By using a lot of mini-baches, the computational time usually increases because the network requires more iterations for each epoch. However, using a full-batch we get faster training while it requires more memory to store all the collocation points for one epoch. In this study, we used a full batch for our training. One main reason is the restriction from the energy loss term where we need to integrate over all the collocation points to compute internal energy.

\item The value of the learning rate is normally between $[1e{-4},~1]$, see \cite{zeiler2012adadelta}. It is also possible to choose an adaptive learning rate based on the value of the loss function, which leads to more intelligent convergence. See also discussions in Appendix A for further possibilities on using a more complex form of learning rate.

\item The number of hidden layers and the number of neurons should be also studied. Having a higher number of layers (i.e. a deeper network) increases the number of neurons and trainable parameters. Choices for these two parameters severely depend on the nature of the problem and to some extent the other parameters which have been mentioned. Therefore, the selected numbers should be studied based on the number of collocation points for each training. We kept the number of neurons in each layer equal, and after several studies, we finally chose $5$ layers with $40$ neurons in each layer. The criteria for preferring this network are: firstly, the final value of the total loss function at the end of the training, and secondly, the absolute error of the output values. Meanwhile, by increasing the number of neurons, overfitting may occur \cite{ying2019overview}. 

\item The number and location of collocation points play an important role. According to discussions in Appendix B and Section~\ref{sec:ther}, adding more collocation points (specially in regions where we expect high gradients of variable) helps to obtain more accurate solutions. This is also similar to the location of nodes and element refinement in FEM. 

\end{itemize}

\subsubsection{A more complex case for heterogeneity}
\begin{figure}[H] 
  \centering
  \includegraphics[width=1.0\linewidth]{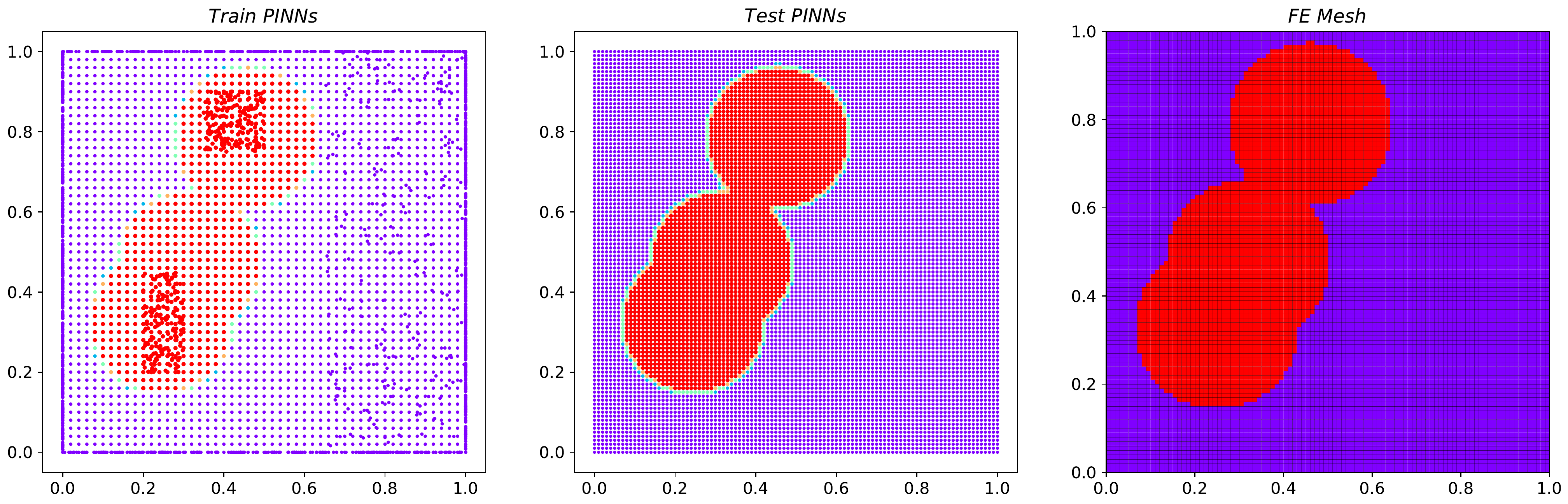}
  \caption{Left: collocation points for the network training. Middle: points at which the trained networks is evaluated. Right: quadrilateral meshes to for FEM. Different colors represent changes in the Young's modulus.}
  \label{fig:geom2_config}
\end{figure}

\begin{figure}[H] 
  \centering
  \includegraphics[width=1.0\linewidth]{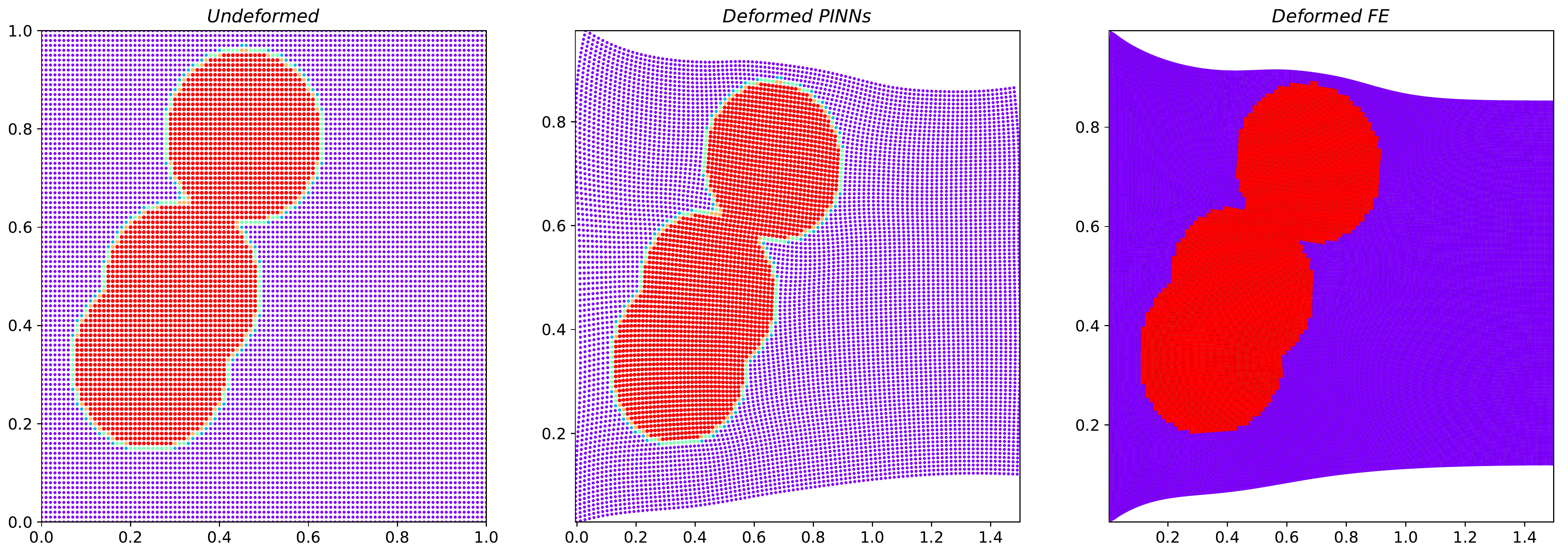}
  \caption{ Undeformed and deformed configuration obtained from the mixed formulation based PINNs and FEM for the mechanical problem considering geometry 2. The results are magnified by a factor of $10$.}
  \label{fig:deform_model2}
\end{figure}

\begin{figure}[H] 
  \centering
  \includegraphics[width=0.95\linewidth]{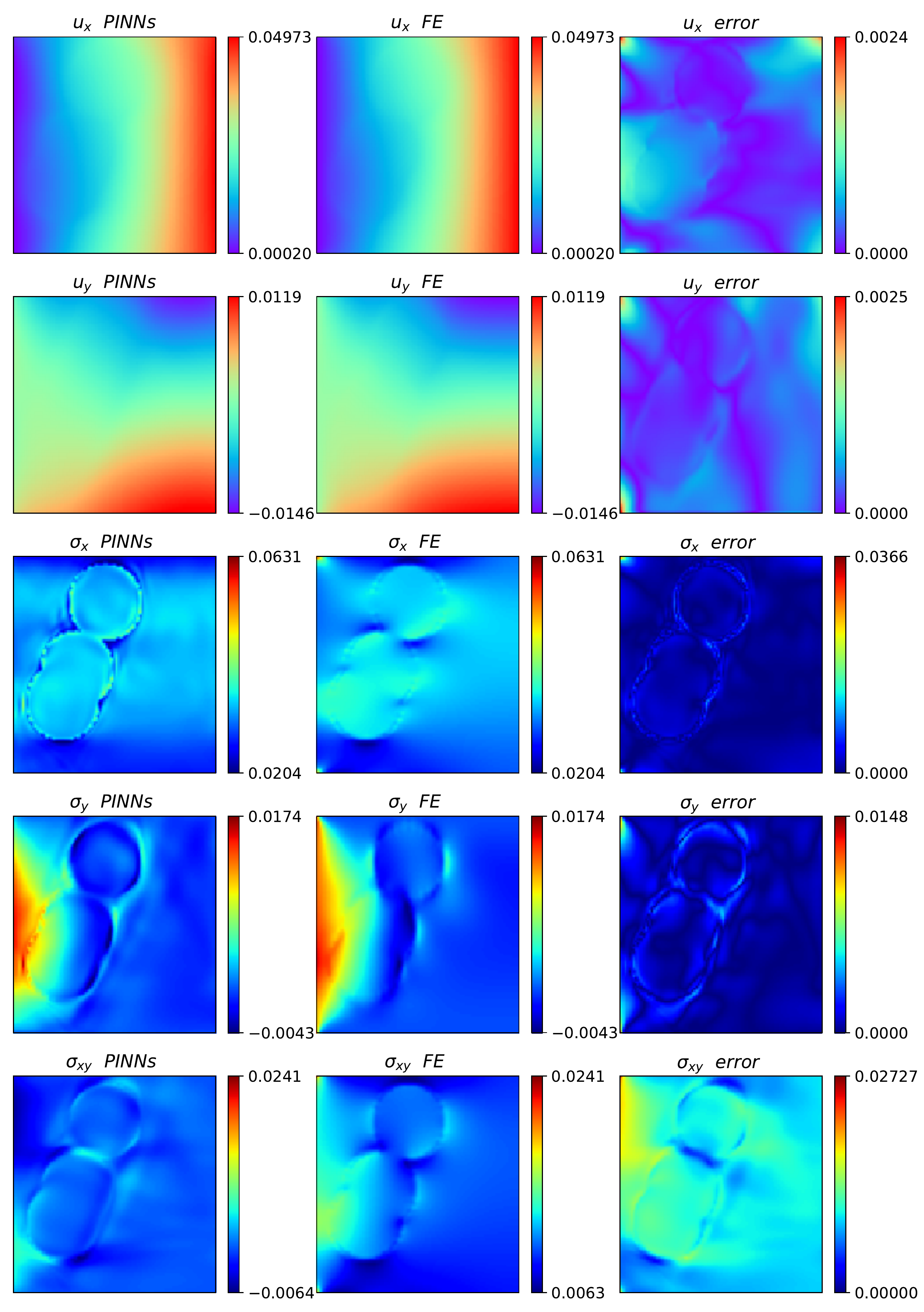}
  \caption{Results for the mechanical problem using geometry 2 and the introduced mixed formulation based PINNs and the FEM.}
  \label{fig:res_geom2}
\end{figure}

To show the applicability of the mixed formulation (model E), we examine a more complex pattern for heterogeneity. For this case, the collocation points are shown on the left-hand side of Fig.~\ref{fig:geom2_config}. 
Here the total number of collocation points is around 5000. On each boundary, we considered 400 points. 
In the middle part of Fig.~\ref{fig:geom2_config}, the points, where the trained network is evaluated are shown. The network is evaluated in a much denser grid ($100 \times 100$ points) where the majority of these points are not located in the set for collocation points. By such selection, we intend to show the power of interpolation in PINNs within the domain of study. The FE approach is also performed on a very fine mesh ($100 \times 100$ elements) shown on the right of Fig.~\ref{fig:geom2_config}. 
The undeformed and deformed shape of the microstructure is shown in Fig.~\ref{fig:deform_model2}. Further results of the study are summarized in Fig.~\ref{fig:res_geom2}. Despite the complex shape of heterogeneity, the network can predict the distribution of deformation and stress components. 
Again, we observe that the main difference is mostly at the corners of the left boundary.

\newpage
\subsection{Other type of BCs and material properties} \label{sec:mix}
\begin{figure}[H] 
  \centering
  \includegraphics[width=1.0\linewidth]{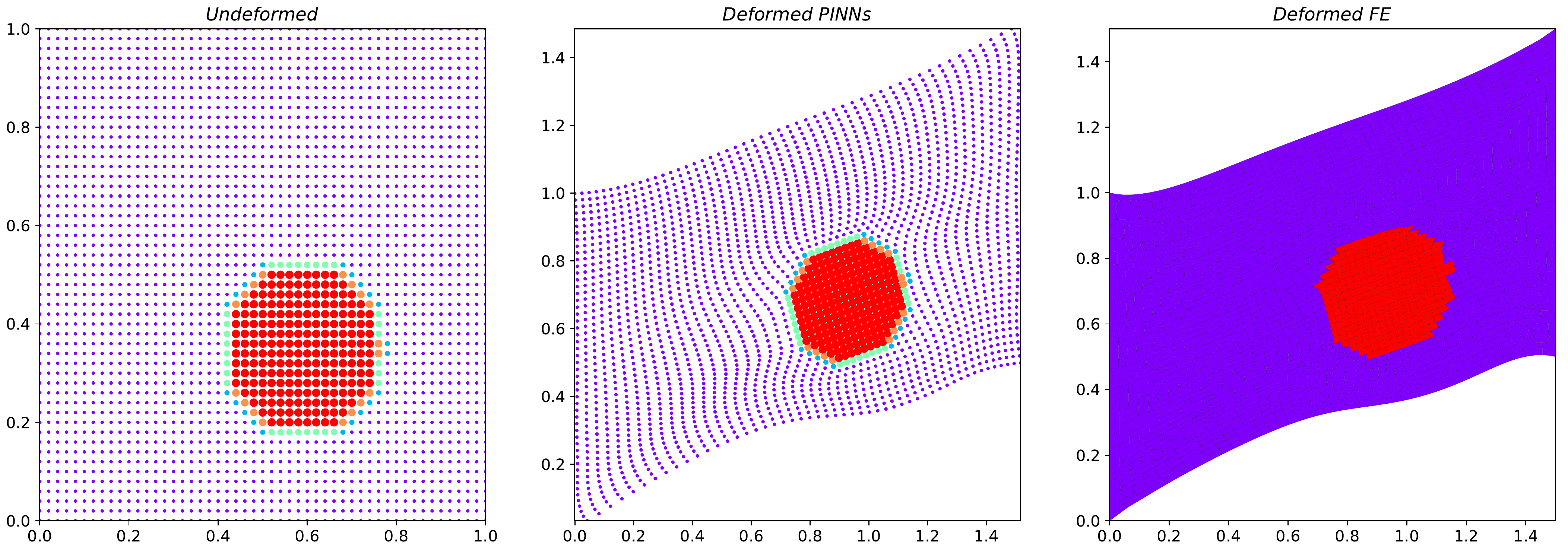}
  \caption{Undeformed and deformed configuration obtained from the introduced mixed formulation based PINNs and FEM considering geometry 1. The results are magnified by a factor of $10$..}
  \label{fig:mixmode_def}
\end{figure}

The mixed formulation based model is trained for simulation of other BCs and material properties. Here, we took the same network described in section~\ref{sec:mech}. On the right edge we applied Dirichlet BCs for both displacement components, i.e. $\bar{u}_x=\bar{u}_y=0.05~\mu$m. The left boundary is completely fixed, and the top and bottom edges are traction free. Moreover, we increase the contrast in the Young's modulus of the phases to a higher number, i.e. $E_{\text{inc}}/E_{\text{mat}} = 10$. The obtained deformed structures utilizing PINNs and FEM are shown in Fig.~\ref{fig:mixmode_def}. 

\begin{figure}[H] 
  \centering
  \includegraphics[width=1.0\linewidth]{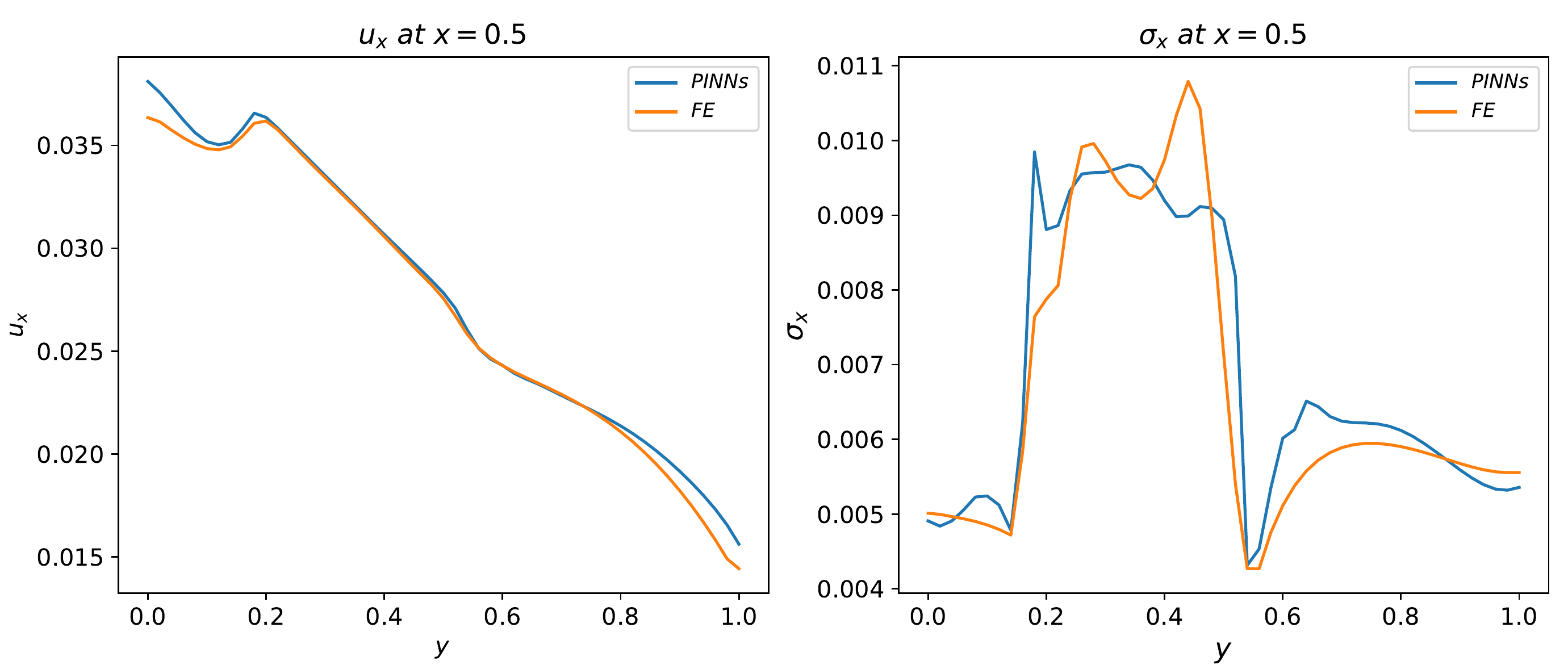}
  \caption{Comparison between the mixed formulation based PINNs and FE along $x=0.5~\mu$m direction for geometry 1.}
  \label{fig:mixmode_sec}
\end{figure}

\begin{figure}[H] 
  \centering
  \includegraphics[width=0.95\linewidth]{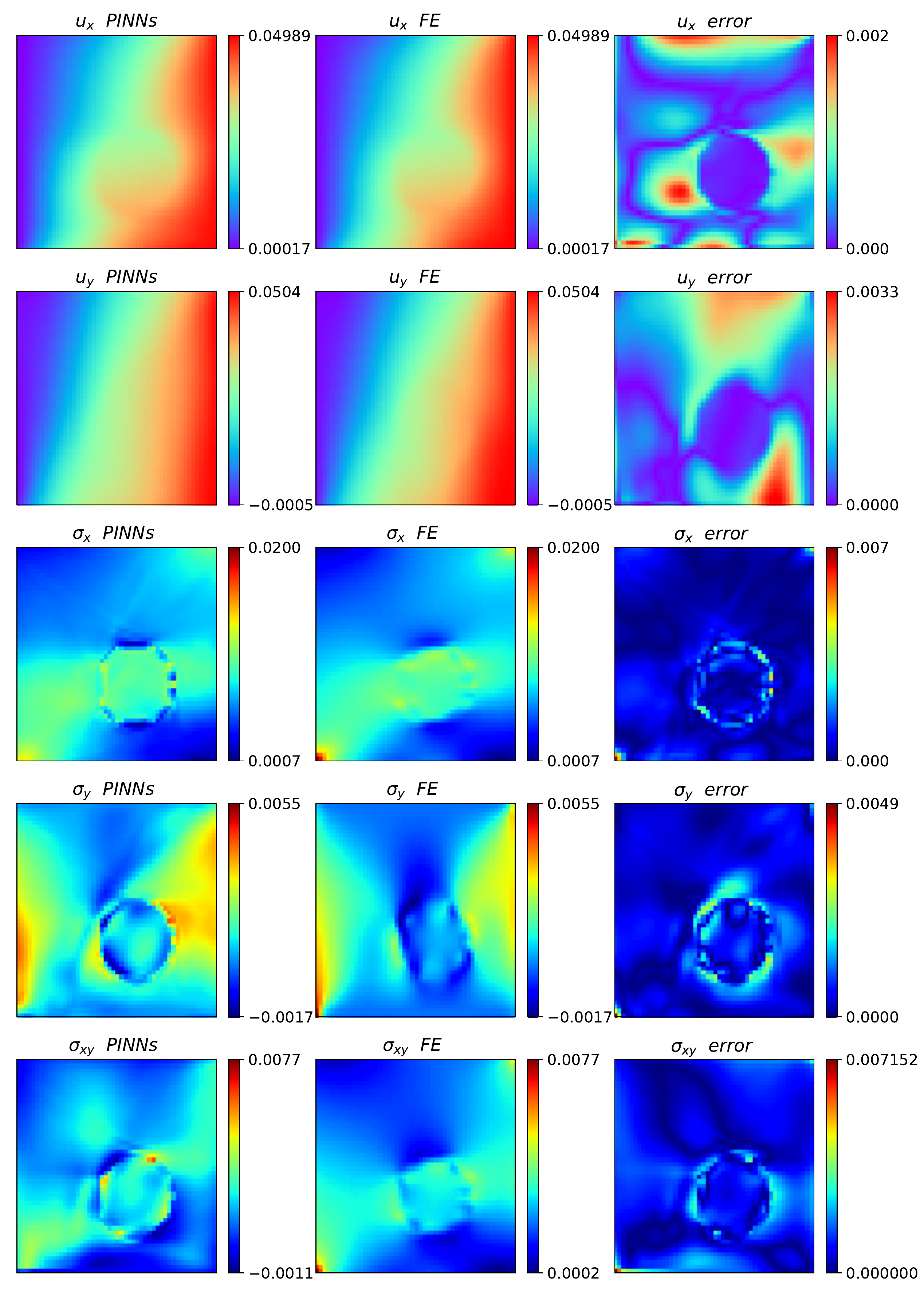}
  \caption{Results for the mechanical problem using new boundary conditions ($\bar{u}_x=\bar{u}_y=0.05~\mu$m) and the new phase contrast ($E_{\text{inc}}/E_{\text{mat}} = 10$).}
  \label{fig:mixmode}
\end{figure}

For a quantitative comparison, displacement as well as stress in $x$ direction at $x=0.5~\mu$m are plotted along the $y$ axis in Fig.~\ref{fig:mixmode_sec}. For the displacement field, we observe a close response between the two approaches except for the error at the boundaries for the mixed formulation based PINNs. For the stress profile, one observes more deviation between the two, but the overall response is well represented. Further comparisons are presented in Fig.~\ref{fig:mixmode}. Here, the values of $\sigma_{xy}$ are higher due to the BCs and in good agreement with FE results. The maximum error for stress components is about $10\%$, and it is localized close to the outer and phase boundaries.

\newpage
\subsection{Diffusion problem}  \label{sec:ther}
For the proposed mixed formulation based PINNs in this case, the output variables are the temperature field $T$ as well as components of the heat flux vector $\bm{q}$. These are approximated via fully connected feed-forward neural networks (FFNN) just like the description provided in Section \ref{sec:mech}.

\begin{figure}[H] 
  \centering
  \includegraphics[width=1.0\linewidth]{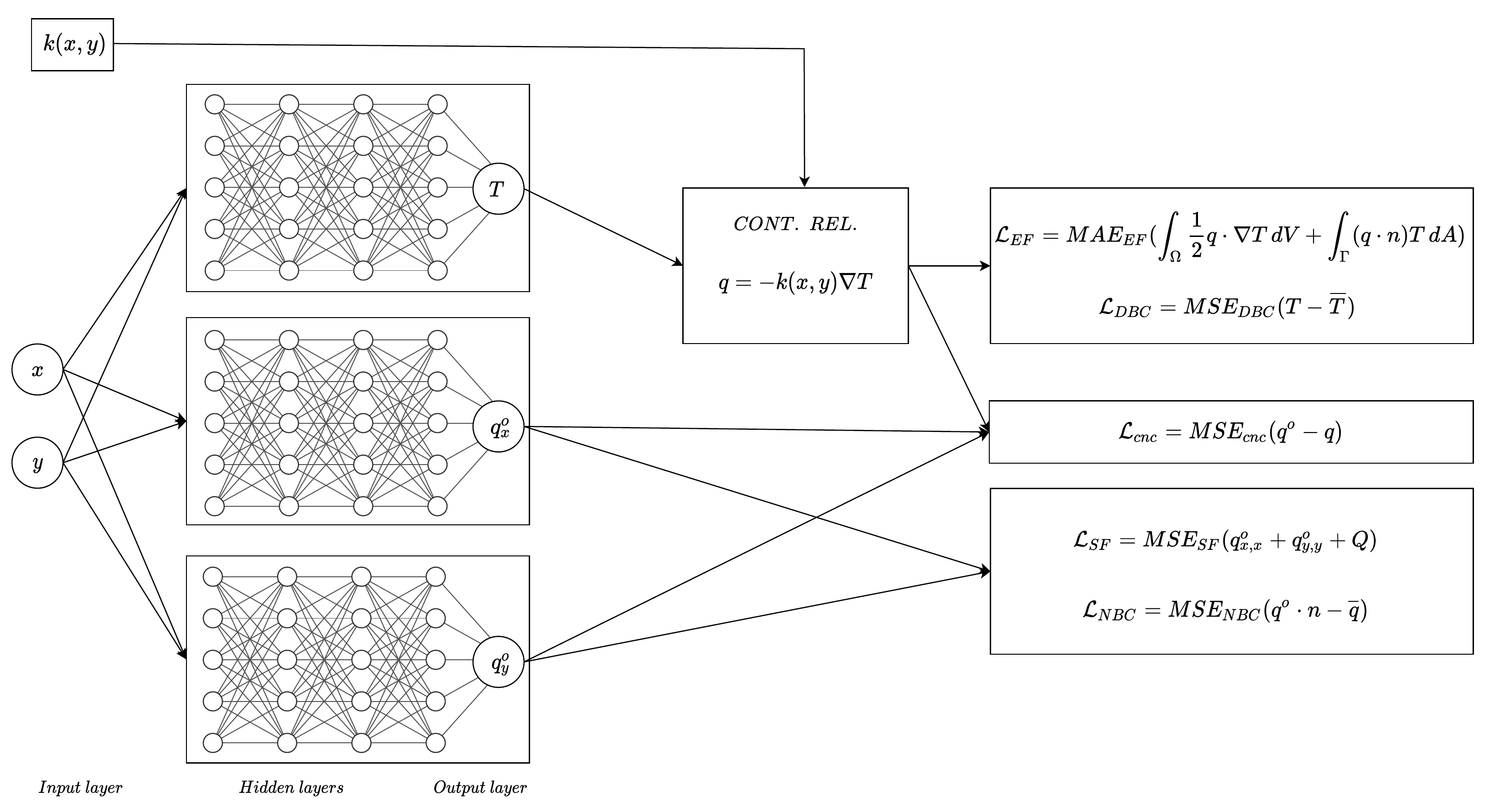}
  \caption{ Architecture and loss functions for the mixed formulation based network in the thermal problem. }
  \label{fig:thermal_arch}
\end{figure}

Similar to before, by denoting each neural network system by $\mathcal{N}$, we write the following solution components for the problem. Note that the proposed network is comparable with model E in Table~\ref{tab:model} where in addition to the primary variable ($T$), its spatial derivative ($\bm{q}^O$) is also set as a network's output with a separated network for each quantity.
\begin{align}
\label{eq:NN_3_T}
    T     &= \mathcal{N}_{T} (\bm{X}; \bm{\theta}), \\
    q^O_x &= \mathcal{N}_{q_x} (\bm{X}; \bm{\theta}), \\
    q^O_y &= \mathcal{N}_{q_y} (\bm{X}; \bm{\theta}).
\end{align}
Based on the described equations for the thermal problem (Eqs.~\ref{Fourier} to \ref{int_Energy_ther}), the network architecture is shown in Fig.~\ref{fig:thermal_arch}. For each network $\mathcal{N}$, the parameters are according to Table~\ref{tab:network}.
The total loss function and its components are defined below (see Eq.~\ref{Totalloss_th} and what follows). The network consists of three main parts. First, we have the loss terms applied to the temperature $T$. These terms are based on the energy form of the problem as well as the Dirichlet BCs ($\mathcal{L}_{EF}$ and $\mathcal{L}_{DBC}$). Next, we add the loss terms applied to the flux vector $\bm{q}$ as well as the Neuman BCs ($\mathcal{L}_{SF}$ and $\mathcal{L}_{NBC}$). Finally, we have the connection loss term $\mathcal{L}_{cnc}$. See also Appendix C, where we further exploited these terms according to the described BVP in Fig.~\ref{fig:BC}. 

Different loss terms are minimized with respect to the free parameters of the network. Again, we start with geometry 1 described in Fig.~\ref{fig:allgeom}. The boundary conditions for the thermal problem are according to the right-hand side of Fig.~\ref{fig:BC}. The position of collocation points for this problem is similar to the mechanical problem and is presented in Fig.~\ref{fig:geom1_config}. A finite element simulation is also performed based on regular quadrilateral meshes shown in Fig.~\ref{fig:geom1_config}. For the FE calculations, standard Q1 elements with bilinear shape functions are utilized \cite{Bayat2020}.
\begin{align}
\label{Totalloss_th}
\mathcal{L} &= \underbrace{\mathcal{L}_{EF} + \mathcal{L}_{DBC}}_{\text{based on $T$}} + \mathcal{L}_{cnc} + \underbrace{\mathcal{L}_{SF}  + \mathcal{L}_{NBC}}_{\text{based on $\bm{q}^o$}}, \\
\label{loss_weak_th}
\mathcal{L}_{EF} &= \text{MAE}_{EF}\left( \int_{\Omega}^{} \frac {1}{2} \bm{q}^T \cdot \nabla T \,dV + \int_{\Gamma}^{} (\bm{q} \cdot \bm{n})~T \,dA \right), \\
\label{loss_DBC_th}
\mathcal{L}_{DBC} &= \text{MSE}_{DBC}\left( T - \overline{T} \right), \\
\label{loss_cnc_th}
\mathcal{L}_{cnc} &= \text{MSE}_{cnc}\left( \bm{q}^o - \bm{q} \right), \\
\label{loss_SF_th}
\mathcal{L}_{SF} &= \text{MSE}_{SF}\left( \text{div}(\bm{q}^o) \right), \\
\label{loss_NBC_th}  
\mathcal{L}_{NBC} &= \text{MSE}_{NBC}\left( \bm{q}^o \cdot \bm{n} - \overline{\bm{q}} \right).
\end{align}
In the above relation, the definition for the mean square error function is according to Eqs.~\ref{eq:MSE} and \ref{eq:type}.

\begin{figure}[H] 
  \centering
  \includegraphics[width=0.85\linewidth]{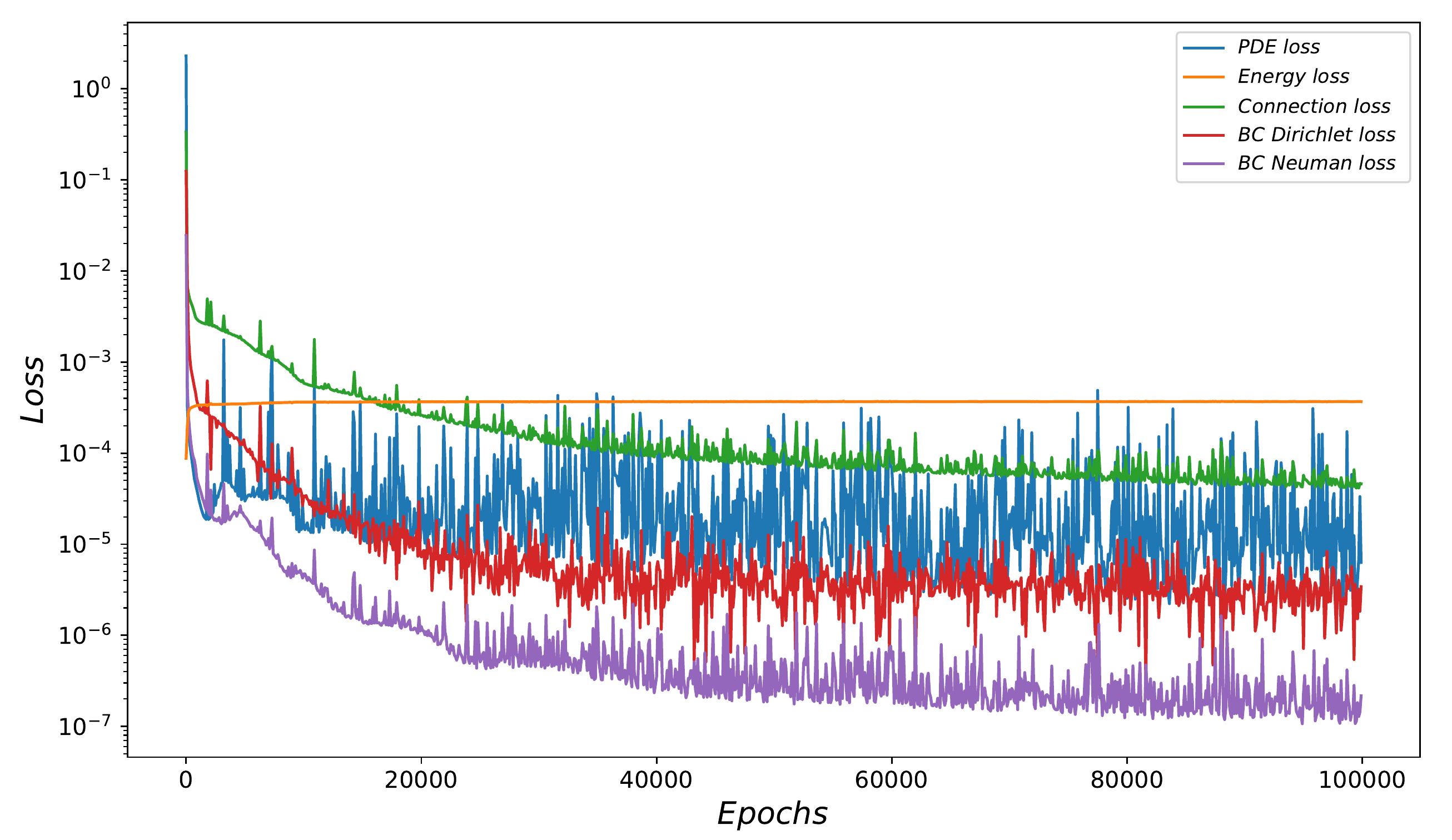}
  \caption{ Individual loss terms versus number of epoch for geometry 1.  }
  \label{fig:all_losses_ther}
\end{figure}

In Fig.~\ref{fig:all_losses_ther}, different loss terms related to the governing equation, its energy form as well as the one related to BCs are plotted versus a number of epochs (iterations for training).
We minimized the overall loss function for the same problem for five different times to make sure that the introduced network is always able to achieve a unique solution. Although not reported but we have observed similar same patterns as in Fig.~\ref{fig:all_losses_ther} by repeating the minimization problem under same conditions.

\begin{figure}[H] 
  \centering
  \includegraphics[width=1.0\linewidth]{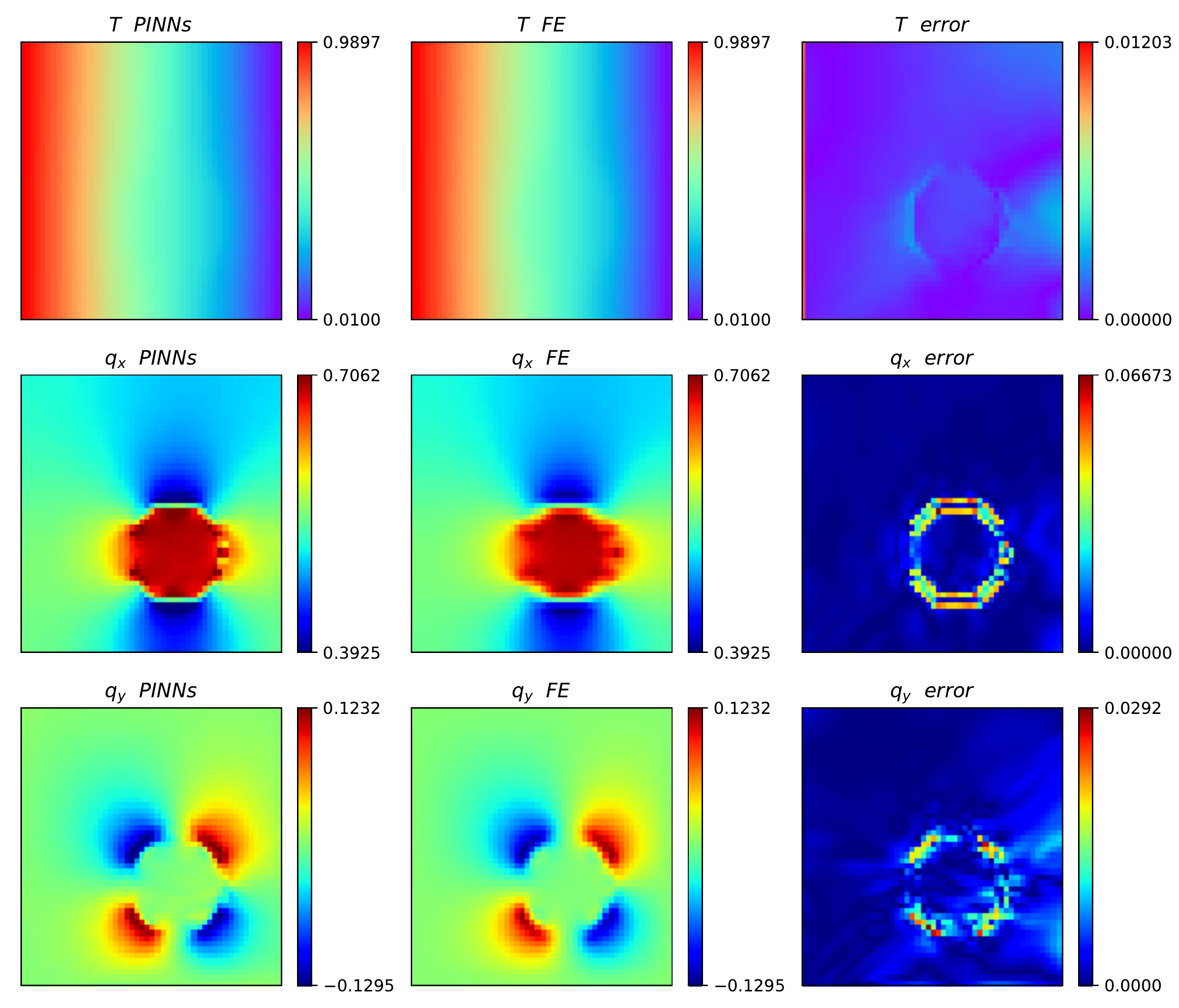}
  \caption{Results for the thermal problem using the mixed formulation based PINNs and the FEM.}
  \label{fig:res_geom1_ther}
\end{figure}

\begin{figure}[H] 
  \centering
  \includegraphics[width=0.9\linewidth]{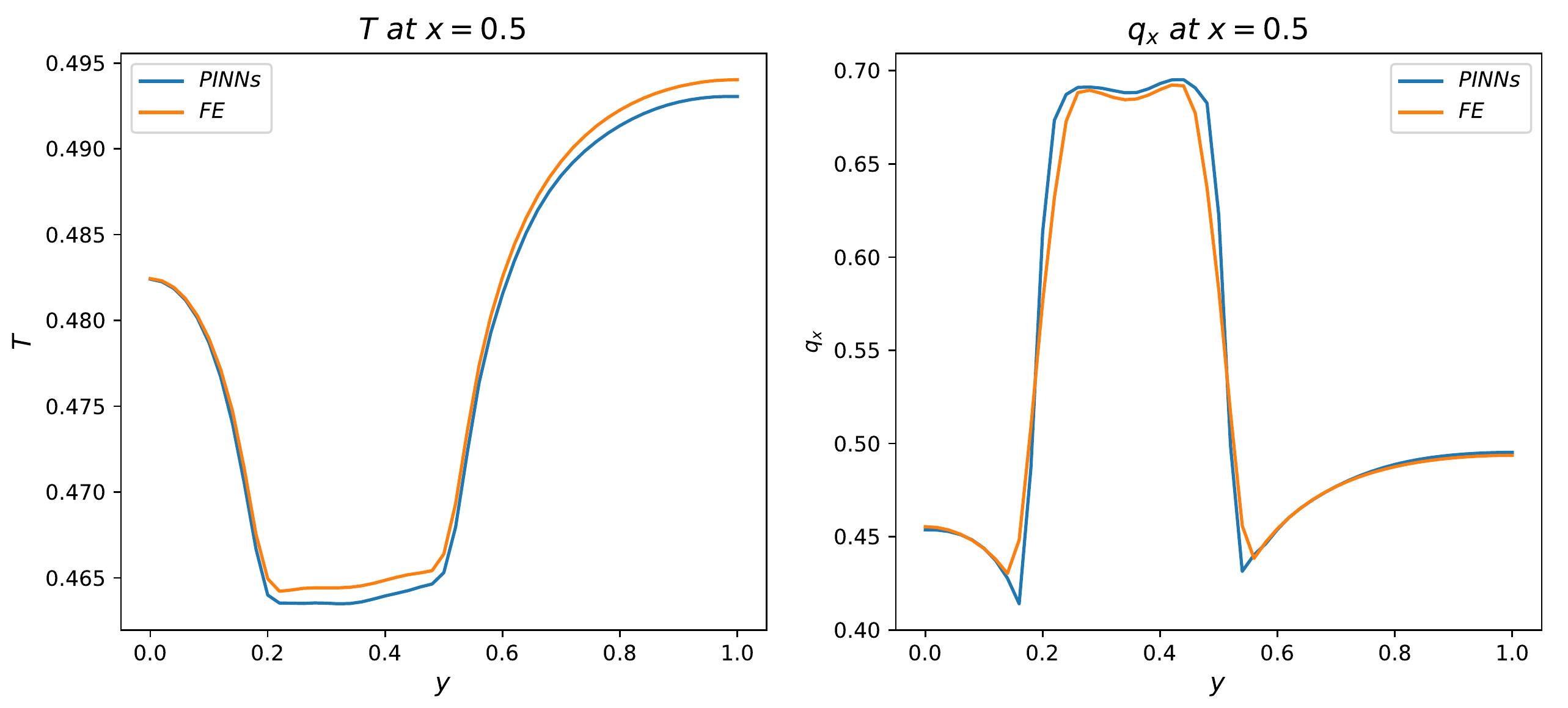}
  \caption{Comparison between the mixed formulation based PINNs and FE along $x=0.5~\mu$m direction.  }
  \label{fig:res_ther_sec}
\end{figure}

\begin{figure}[H] 
  \centering
  \includegraphics[width=1.0\linewidth]{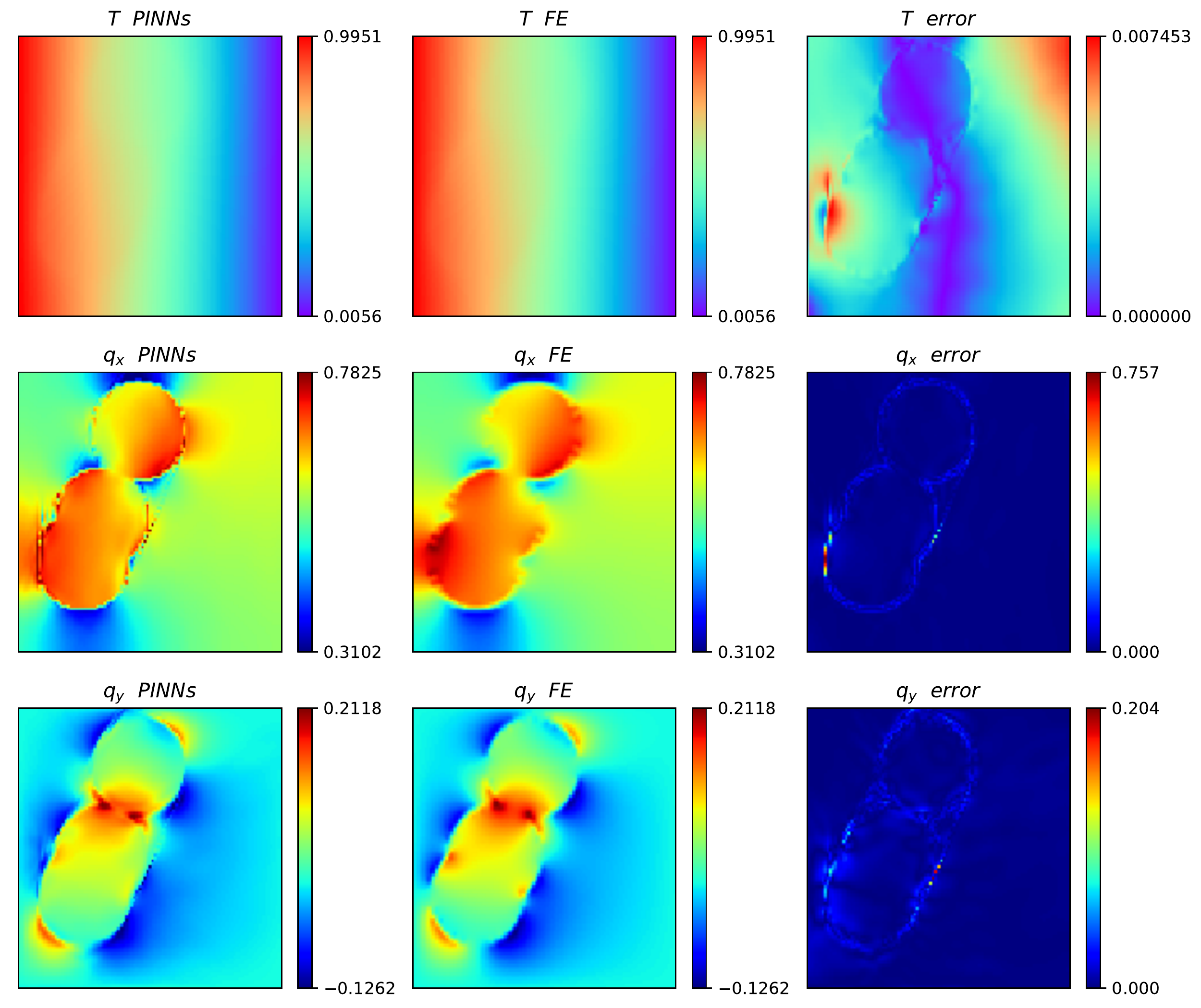}
  \caption{Obtained results from the mixed formulation based PINNs and the FEM for the thermal problem.}
  \label{fig:Results_geom2_ther}
\end{figure}

The obtained results from PINNs and FEM are reported in Fig.~\ref{fig:res_geom1_ther} wherein the first row is the predicted temperature field and in the last two, the components of the heat flux vector are presented. In the first, second, and third columns, the results from PINNs, FE, and the difference between these two methodologies are presented, respectively. 
For all the output values, we observe a close agreement between the network predictions and those from the FE method. The difference is bold in the area where we have the sharp phase transition (border of heterogeneity). The maximum relative difference is about $12 \%$ and $9.6 \%$ for temperature and flux values, respectively. The average relative difference value is about $0.5 \%$ and $0.6 \%$ for temperature and flux values, respectively.
Moreover, the results for predicted $T$ and $q_{x}$ along the cross-sections $x=0.5~\mu$m are shown in Fig.~\ref{fig:res_ther_sec}.
The mixed formulation based network for the thermal problem is applied to examine more challenging cases. For this purpose, we use geometry 2, where we have a random complex heterogeneity. The collocation points are shown on the left-hand side of Fig.~\ref{fig:geom2_config}. The middle part of the figure depicts the points where the trained network is evaluated. Note that the network is evaluated in a much denser grid ($100 \times 100$ points). In other words, many of these points are not located in the set for minimization of the loss function (i.e., collocation points). The FE approach is also performed on a fine mesh ($100 \times 100$ elements) shown on the right of Fig.~\ref{fig:geom2_config}. The results of the study are summarized in Fig.~\ref{fig:Results_geom2_ther}. Despite the complex shape of heterogeneity, the network can predict the distribution of deformation and stress components. The main difference is localized at the specific places in phase boundaries.

One possible remedy to reduce this error is to consider more collocation points (see Appendix B). As shown on the left-hand side of Fig.~\ref{fig:refine}, we add about 800 collocation points in the vicinity of the interphase region where the highest error is observed. Loss functions of the same network are minimized for the new set of collocation points. According to Fig.~\ref{fig:refine}, we observe a significant reduction in the difference between the mixed formulation based PINN and FE results using more collocation points.
\begin{figure}[H] 
  \centering
  \includegraphics[width=1.0\linewidth]{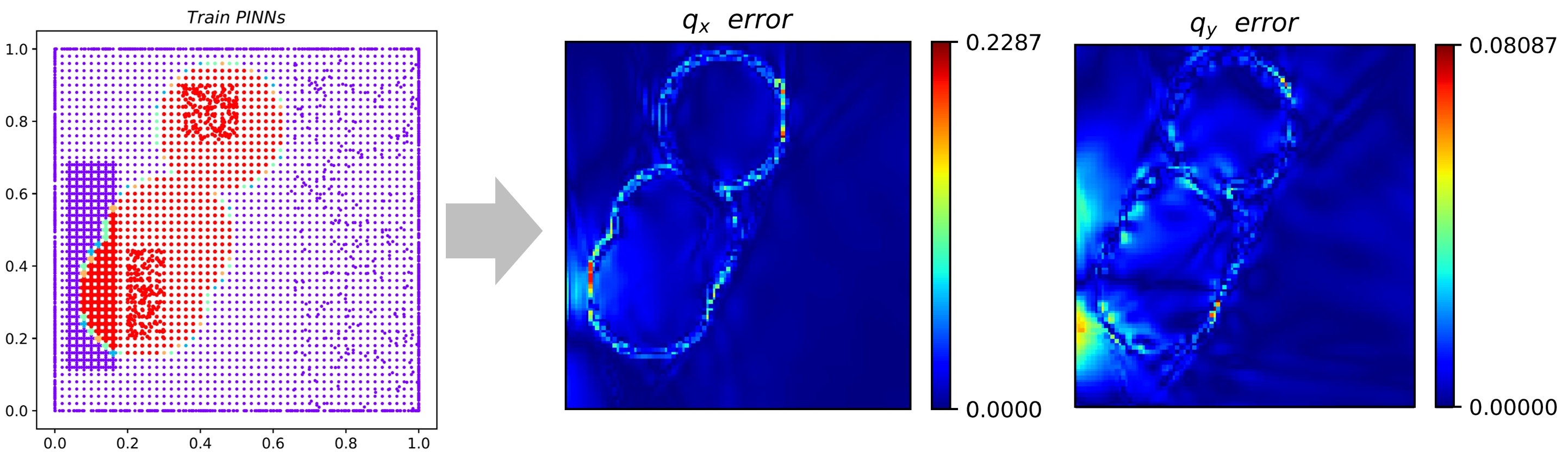}
  \caption{Difference between the mixed formulation based PINN and FE after refinement of collocation points at the interface region. The results obtained from PINN are (by two orders for $q_y$) improved after minimizing the network using the new set of collocation points shown on the left-hand side.}
  \label{fig:refine}
\end{figure}


\section{Conclusion and outlooks}
The potential of physics-informed neural networks for finding the solution to a given boundary value problem in a heterogeneous domain is studied. Heterogeneity is an important aspect related to the material microstructure. It offers a lot of flexibility for material design, and therefore accurate and fast prediction of material behavior at the microscale is essential. By combining several well-known ideas from the literature on PINN besides FE technology, we extend the available PINN models to a mixed formulation based version. Here the idea is to introduce the spatial gradient of the primary variable as an additional output for the network. Furthermore, a separated network is addressed for each output. Then, the output quantities are connected and constrained based on the governing equation (PDE) and its energy form (obtained from the weak form). 
Two problems governed by mechanical equilibrium plus a steady-state diffusion equation are investigated.
The results from DL are compared against the solution of the standard FEM. Early studies show that the proposed approach predicts the solution without having any initial data. The current work also summarizes overriding knowledge for people in computational science to utilize DL methods in their studies. 

Despite many valuable contributions \cite{RAISSI2019, Guo2022, Henkes2022}, further studies are needed to fully understand the role of the networks (hyper) parameter and their relation to the final obtained solution (e.g., layer depth, learning rate, location, and distribution of collocation points, etc.). Moreover, one of the main advantages of PINNs is their simple implementation. Integrating new boundary conditions and complex geometries or investigating new governing equations are straight forward. The main drawback for PINNs at the current moment is the computational time. For instance, for the problems discussed in this work one needs in order of few hours to minimize the loss function and find the solution, whereas the same problem took less than a few minutes using FEM (upon having the developed codes and models). Certainly, the optimizer or design of the network for DL should be improved to speed up this process. As examples of developments in this direction, one can mention utilizing other minimizers based on Broyden–Fletcher–Goldfarb–Shanno (BFGS) methods \cite{Henkes2022}, enhancing the loss functions with additional gradient terms \cite{YU2022114823} or breaking the given domain to sub-domains and training a separate network for each one \cite{JAGTAP2020113028}.

\begin{figure}[H] 
  \centering
  \includegraphics[width=0.95\linewidth]{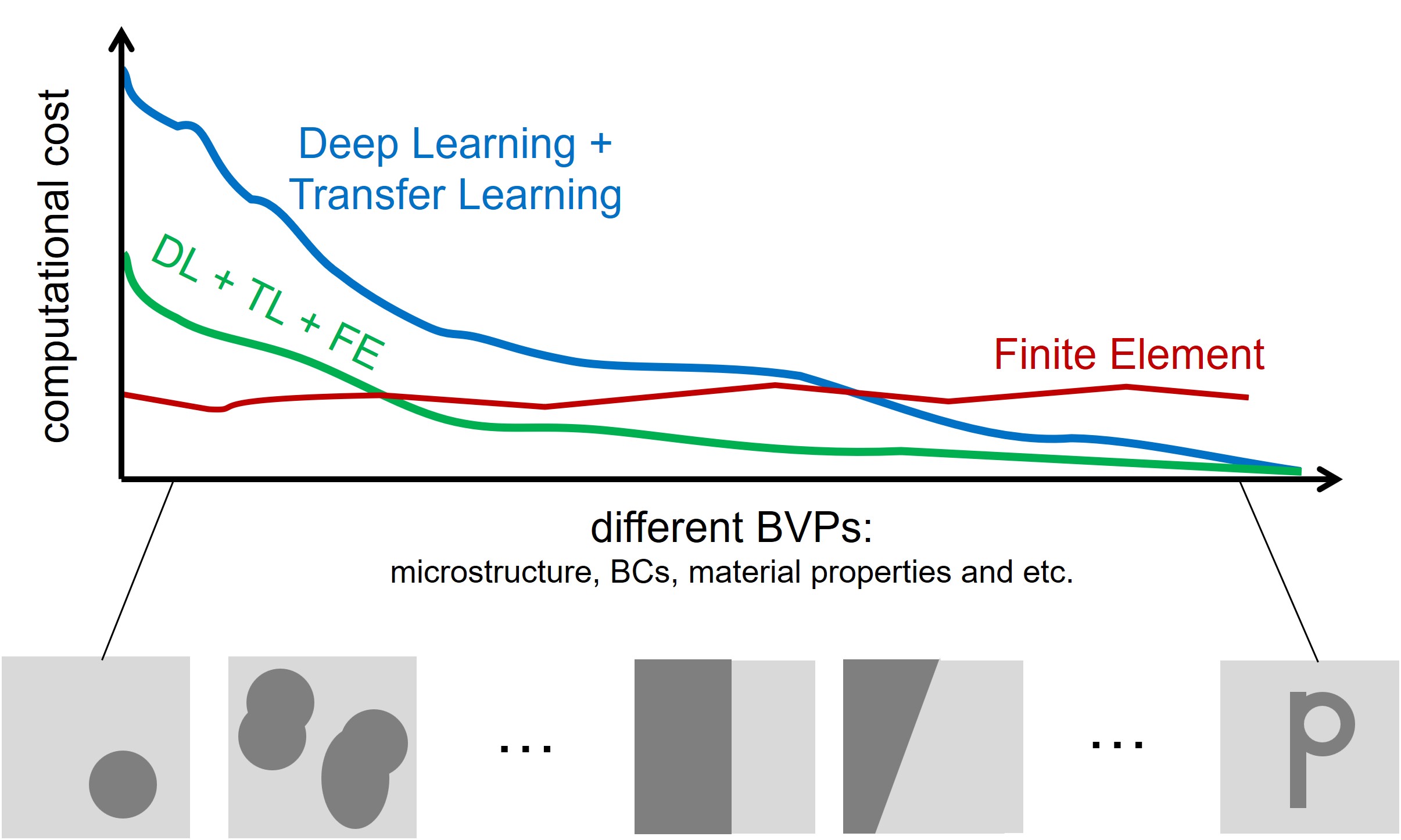}
  \caption{A comparison between available solver such as finite element (FE) and upcoming methodologies such as deep learning (DL). At the moment the computational cost of DL is higher than FE. This cost can go to zero by combining the two methods as well as utilizing transfer learning (TL) in the future developments. }
  \label{fig:presp}
\end{figure}

One should note that the DL methods can be trained for different BVPs and various loading conditions and material properties (i.e., all the interesting parameters one requires in each engineering task). For example the distribution of the stiffness matrix or different type of boundary conditions can be an additional input for the network. As a result, the computational cost of these methods will decay as they get trained by realizing the new systems. See schematic representation of this latter discussion in Fig.~\ref{fig:presp}. The last point, in the long-term, has the potential to solve two main issues from computational science:  computational cost and accuracy simultaneously. Furthermore, the available data from different computational strategies in various domains shall not be wasted and can be used (recycled) for training the ultimate network. Please note that there is also another strategy where we feed the network with a sufficient amount of data or situation (case studies) at once and then train the network. See for example \cite{Krishna2021, mianroodi2021teaching, mianroodi2022lossless, Yangeabd7416, wang2021train}. 

This work focused on isotropic elastic material in a 2D setting. Therefore in future developments, one should investigate this further for geometrical and material nonlinear settings in 3D. Moreover, it is interesting to explore the idea of solving multiphysics problems where we also have system evolution in time.

\color{black}

\section{Appendix A: ideas to enhance the learning rate based on the Newton-Raphson method}
For the training, we intend to minimize the total loss function $\mathcal{L}$ with respect to the network parameter $\bm{\theta}$ (i.e. $\min_{\bm{\theta}} \mathcal{L}(\bm{X}; \bm{\theta})$). As discussed in section~\ref{sec:mech}, through an iterative process, one can reach the (global) minimum point using gradient decent method:
\begin{align}
\label{iteration_app}
\bm{\theta}_{i+1} = \bm{\theta}_{i} - \alpha  \nabla_{\bm{\theta}}\mathcal{L}(\bm{\theta}_{i}).
\end{align}
On the other hand, at the minimum point one can argue that we have 
\begin{align}
\label{min_point}
\nabla_{\bm{\theta}}\mathcal{L}(\bm{\theta}^*) = \dfrac{\partial \mathcal{L}}{\partial \bm{\theta}}\bigg|_{\bm{\theta}^*}=\bm{0}=\bm{L}(\bm{\theta}^*).
\end{align}
In other words, the minimization process can be interpreted as a root-finding problem and the solution is where $\bm{L}$ goes towards zero. For this problem, we apply the standard Newton-Raphson method. Therefore, one can write
\begin{align}
\label{newton}
\bm{\theta}_{i+1} = \bm{\theta}_{i} - 
\left(\dfrac{\partial \bm{L}}{\partial \bm{\theta}}\right)^{-1}\bigg|_{\bm{\theta}_{i}} \bm{L}(\bm{\theta}_{i}).
\end{align}
According to our definition we have $\bm{L}(\bm{\theta})=\nabla_{\bm{\theta}}\mathcal{L}(\bm{\theta})$. Substituting the latter expression into Eq.~\ref{newton}, we conclude that
\begin{align}
\label{newton2}
\bm{\theta}_{i+1} = \bm{\theta}_{i} - 
\underbrace{\left(\dfrac{\partial^2 \mathcal{L}}{\partial \bm{\theta}^2}\right)^{-1}\bigg|_{\bm{\theta}_{i}}}_{\bm{\alpha}_i} \nabla_{\bm{\theta}}\mathcal{L}(\bm{\theta}_{i}).
\end{align}
Comparing Eq.~\ref{iteration_app} and Eq.~\ref{newton2}, one can conclude that for a faster convergence to the minimum point, the learning rate can take a tonsorial form $\bm{\alpha}$. The latter suggests that for going in direction of the minimum point, one can consider some weightings to different components of the vector $\nabla_{\bm{\theta}}\mathcal{L}(\bm{\theta}_{i})$ (instead of a scalar number). On the other hand, the calculation of the second derivative of the loss function and its inversion might not be trivial and gets costly. Therefore, a proper approximation of this tensor quantity can be extremely beneficial for future studies. See also other algorithms such as BFGS methods for optimization and training \cite{Henkes2022}. \\ \\

\newpage
\section{Appendix B: Influence of the derivation order, number of collocation points, and number of epochs on the obtained solution}
We consider the following function $y(x)$ as the desired solution:
\begin{align}
\label{sol}
y(x) = \text{sin}(x) + 0.1~x + 0.1.
\end{align}
To construct our differential equations, we take the first and the second derivative of the above relation concerning the parameter $x$. Therefore, we write the following 1st and 2nd order ordinary differential equations (ODEs) subjected to the following boundary conditions:
\begin{align}
\label{def_1}
y'  &= \dfrac{\text{d}y}{\text{d}x} = \text{cos}(x) + 0.1,~~~y(0)=0.1,~~~y(10)=0.556, \\
\label{def_2}
y'' &= \dfrac{\text{d}^2y}{\text{d}x^2} = -\text{sin}(x),~~~~~~~y(0)=0.1,~~~y(10)=0.556.
\end{align}
The idea is to obtain (reproduce) the function $y(x)$, by feeding the above-mentioned ODEs to two separate neural networks with the same architecture and hyper parameters. As a result, one can study the influence of derivation order on the performance of the deep learning algorithm which is used to find the solution $y(x)$. 
Therefore, we define the following loss functions based on the above ODEs and their relative BCs:
\begin{align}
\label{loss_ode_1}
\mathcal{L}_{o1} &= \text{MSE}\left(\dfrac{\text{d}y_{o1}}{\text{d}x} - \text{cos}(x) - 0.1 \right) + \text{MSE}\left(y_{o1}(0)-0.1\right) + \text{MSE}\left(y_{o1}(10)-0.556\right), \\
\label{loss_ode_2}
\mathcal{L}_{o2} &= \text{MSE}\left(\dfrac{\text{d}^2y_{o2}}{\text{d}x^2} + \text{sin}(x)\right) + \text{MSE}\left(y_{o2}(0)-0.1\right) + \text{MSE}\left(y_{o2}(10)-0.556\right).
\end{align}
For having a fair comparison, the above loss functions are minimized by utilizing the same network, which has $4$ layers and $20$ neurons in each layer. Moreover, we used \emph{tanh} as an activation function for all the neurons, and we kept the learning rate as 0.001 in the Adam optimizer. 
In the below figure, results obtained from these two networks ($y_{o1}$ and $y_{o2}$) as well the as the expected solution ($y_{true}(x)=y(x)$) are plotted for a different number of collocation points and epochs. The gray area represents the training zone where we have distributed the collocation points equally distanced from 0 to 10. 
The obtained solutions are plotted further (up to 15) to show the extrapolation behavior. On the left-hand side, we observe a poor prediction since we used very few collocation points and a rather low number of iterations (epochs). The predictions improved by adding more collocation points and/or increasing the number of epochs. This study clearly shows that by having a lower degree of derivation, the predictions become more accurate (compare orange and blue curves). If we increase the number of collocation points and epochs, the results of two given ODEs converge to the exact solution. 
\begin{figure}[H] 
  \centering
  \includegraphics[width=1.0\linewidth]{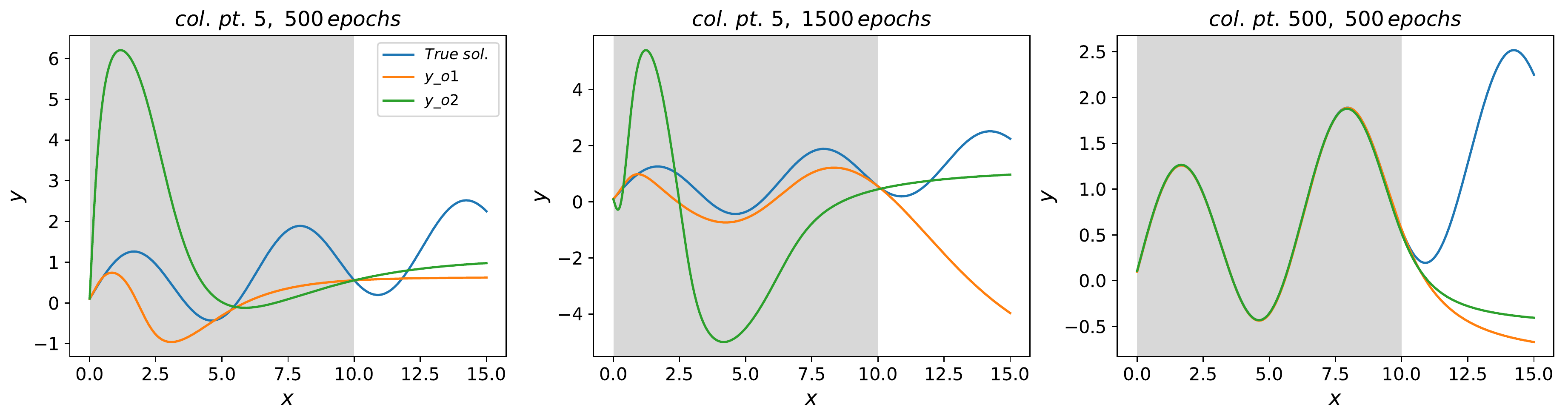}
  \caption{Comparison between the solution obtained from first and second order ODEs. }
  \label{fig:odesin}
\end{figure}

\begin{figure}[H] 
  \centering
  \includegraphics[width=0.8\linewidth]{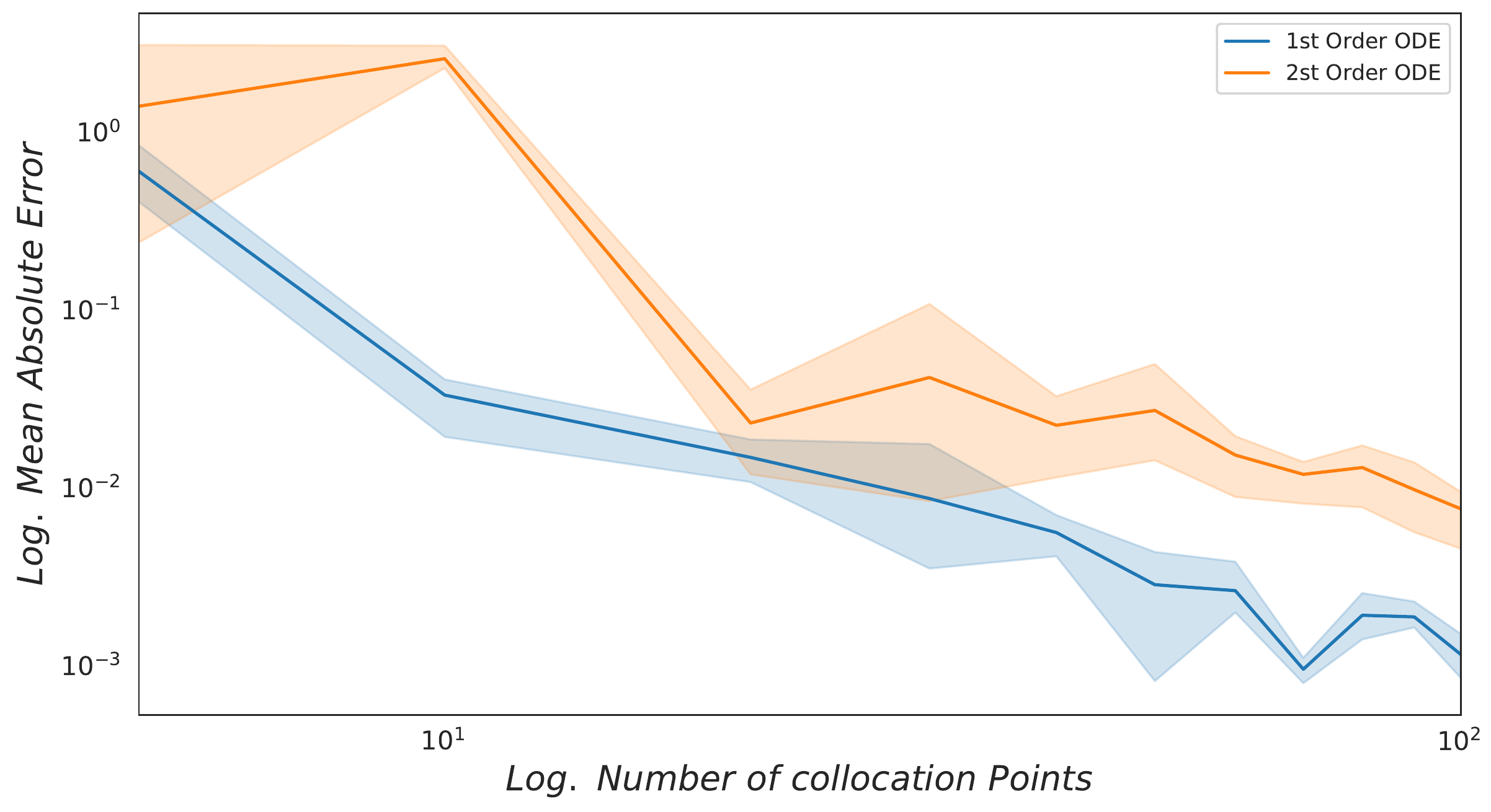}
  \caption{ MAE by increasing the number of collocation points for solution of the 1st and 2nd order ODE. }
  \label{fig:doe_sen_conf}
\end{figure}

To demonstrate the influence of the number of collocation points, we compare the convergence of both forms of the first and second-order ODEs based on the different number of collocation points. The collocation points are changed from 10 to 100 and we kept the number of epochs at 1500 for all the calculations. Since the initial values of trainable parameters could affect the final solution, we repeated the training three times. After each training, we calculated the mean of the absolute difference between the predicted values to the true solutions. The results are reported in Fig.~\ref{fig:doe_sen_conf}. First, we observe that by increasing the number of collocation points, the solution from both types of ODEs convergences. Second, the MAE of first-order ODEs is always lower than second-order ODEs. It is worth to mention that for the 1st order ODE, one needs only one boundary condition while we provide the network with two (see Eq.~\ref{loss_ode_1}). This point might be another explanation for lower errors in the solution of the 1st order ODE. Finally, The randomness or stability of both training is reduced by using more collocation points. After a certain number of collocation points, the standard deviation of the prediction in first-order ODEs is negligible. 



\newpage
\section{Appendix C: Details on the calculation of the loss functions}
For the BVP in Fig.~\ref{fig:BC}, we further expand the loss functions introduced in Eqs.~\ref{Totalloss}-\ref{loss_NBCx}.
First, components of the strain tensor are computed based on the kinematics relation and by utilizing the components of the deformation vector from the network output: 
\begin{align}
\mathcal{N}_{\varepsilon_x}(\bm{X};\bm{\theta}) &= \partial_x \mathcal{N}_{u_{x}},\\
\mathcal{N}_{\varepsilon_y}(\bm{X};\bm{\theta}) &= \partial_y \mathcal{N}_{u_{y}},\\
\mathcal{N}_{\varepsilon_{xy}}(\bm{X};\bm{\theta}) &= \dfrac{1}{2}\left(\partial_{y} \mathcal{N}_{u_{x}} + \partial_{x} \mathcal{N}_{u_{y}} \right).
\end{align}
Next, by employing the constitutive relations we have the components of the stress tensor as 
\begin{align}
\begin{split}
& \mathcal{N}_{\sigma_x}(\bm{X};\bm{\theta}) = \dfrac{E(\bm{X})}{(1+\nu(\bm{X}))(1-2\nu(\bm{X}))}\left((1-\nu(\bm{X}))\mathcal{N}_{\varepsilon_x}+\nu(\bm{X})  \mathcal{N}_{\varepsilon_y} \right) , \\
& \mathcal{N}_{\sigma_y}(\bm{X};\bm{\theta}) = \dfrac{E(\bm{X})}{(1+\nu(\bm{X}))(1-2\nu(\bm{X}))}\left(\nu(\bm{X})\mathcal{N}_{\varepsilon_x}+(1-\nu(\bm{X}))  \mathcal{N}_{\varepsilon_y} \right), \\
& \mathcal{N}_{\sigma_{xy}}(\bm{X};\bm{\theta}) = \dfrac{E(\bm{X})}{2(1+\nu(\bm{X}))}\mathcal{N}_{\varepsilon_{xy}}.
\end{split}
\end{align}
Please note that, for simplicity, the dependency of neural network $\mathcal{N}(\bm{X};\bm{\theta})$ on $\bm{X}$ (input parameters), and $\bm{\theta}$ (network trainabale parameters) is not shown on the right-hand side of the above equations and in the loss terms. The similar trend is utilized also for the thermal part. 
Finally, the loss functions are further expanded as below
\begin{align}
\label{loss_weak_exp} 
\mathcal{L}_{EF} &= \left| - \dfrac{1}{2 N_{b}}\sum_{\bm{X}\in \Omega} 
\left( 
\mathcal{N}_{\sigma_x} \mathcal{N}_{\varepsilon_x} + \mathcal{N}_{\sigma_y} \mathcal{N}_{\varepsilon_y} +
\mathcal{N}_{\sigma_{xy}} \mathcal{N}_{\varepsilon_{xy}} \right)   \,+\,
\dfrac{1}{N^R_{NBC}}\sum_{\bm{X}\in \Gamma^R_N}  \, 
\left(\mathcal{N}_{\sigma_x} \mathcal{N}_{u_x}\right)
\right|, \\
\label{loss_DBC_exp} 
\mathcal{L}_{DBC}  &= \dfrac{1}{N^L_{DBC}} \sum_{\bm{X}\in \Gamma^L_D} \left( \left(\mathcal{N}_{u_{x}}-0\right)^2  + 
\left(\mathcal{N}_{u_{y}}-0\right)^2 \right) + 
\dfrac{1}{N^R_{DBC}} \sum_{\bm{X}\in \Gamma^R_D} \left(\mathcal{N}_{u_{x}}-0.05\right)^2, \\
\label{loss_cnc_exp} 
\mathcal{L}_{cnc} &= \dfrac{1}{N_{b}} \sum_{\bm{X}\in \Omega} \left( \left(\mathcal{N}_{\sigma_{x}^o} - \mathcal{N}_{\sigma_{x}}\right)^2  + 
\left(\mathcal{N}_{\sigma_{y}^o} - \mathcal{N}_{\sigma_{y}}\right)^2  +
 \left(\mathcal{N}_{\sigma_{xy}^o} - \mathcal{N}_{\sigma_{xy}}\right)^2
\right), \\
\label{loss_SF_exp} 
\mathcal{L}_{SF} &= \dfrac{1}{N_{b}} \sum_{\bm{X}\in \Omega} \left( \left(\partial_x \mathcal{N}_{\sigma_{x}^o}\,+\, \partial_y \mathcal{N}_{\sigma_{xy}^o}\right)^2  + 
\left(\partial_y \mathcal{N}_{\sigma_{y}^o}\,+\, \partial_x \mathcal{N}_{\sigma^o_{xy}}\right)^2 
\right), \\
\label{loss_NBCx_exp} 
\mathcal{L}_{NBC} &= 
\dfrac{1}{N^T_{NBC}} \sum_{\bm{X}\in \Gamma^T_N} \left( \left(\mathcal{N}_{\sigma^o_{y}}-0\right)^2  + 
\left(\mathcal{N}_{\sigma^o_{xy}}-0\right)^2 \right)  
 \\
&+ \dfrac{1}{N^B_{NBC}} \sum_{\bm{X}\in \Gamma^B_N} \left( \left(\mathcal{N}_{\sigma^o_{y}}-0\right)^2  + 
\left(\mathcal{N}_{\sigma^o_{xy}}-0\right)^2 \right)
 \\
&+ \dfrac{1}{N^R_{NBC}} \sum_{\bm{X}\in \Gamma^R_N} \left(\mathcal{N}_{\sigma^o_{xy}}-0\right)^2 .
\end{align}
In the above relations $\Gamma^L$, $\Gamma^R$, $\Gamma^T$ and $\Gamma^B$ denote the boundaries (points) at left, right, top and bottom boundary. The sub-index \emph{DBC} and \emph{NBC} stand for Dirichlet and Neumann boundary conditions. 

Note that in the above relations, we simply omit the terms which are zero. For example, having traction-free boundaries on top and bottom and having fixed displacement on the left edge resulted in further simplifications.

For the BVP in Fig.~\ref{fig:BC}, we further expand the loss functions introduced in Eqs.~\ref{Totalloss_th}-\ref{loss_NBC_th}.
First, components of the temperature gradient vector are computed based on the the temperature value from the network output: 
\begin{align}
\mathcal{N}_{\nabla_xT}(\bm{X};\bm{\theta}) &= \partial_x\mathcal{N}_T,\\
\mathcal{N}_{\nabla_yT}(\bm{X};\bm{\theta}) &= \partial_y\mathcal{N}_T.
\end{align} 
Next, we compute the components of the flux tensor according to
\begin{align}
\mathcal{N}_{q_x}(\bm{X};\bm{\theta}) &= -k(\bm{X})\mathcal{N}_{\nabla_xT},\\
\mathcal{N}_{q_y}(\bm{X};\bm{\theta}) &= -k(\bm{X})\mathcal{N}_{\nabla_yT}.
\end{align}
Finally, the loss functions are further expanded as below
\begin{align}
\label{loss_EF_T}
\mathcal{L}_{EF} &= 
\left| 
\dfrac{1}{2 N_b}\sum_{\bm{X}\in \Omega}
\left( 
\mathcal{N}_{q_x} \mathcal{N}_{\nabla_x T} + \mathcal{N}_{q_y} \mathcal{N}_{\nabla_y T} \right) + 
\dfrac{1}{N_{NBC}^L} \sum_{\bm{X}\in \Gamma_N^L}
\left(\mathcal{N}_{q_x} \mathcal{N}_T \right)
\right|, \\
\mathcal{L}_{DBC} &= \dfrac{1}{N^L_{DBC}} \sum_{\bm{X}\in \Gamma^L_D} \left(\mathcal{N}_{T}-1\right)^2 + 
\dfrac{1}{N^R_{DBC}} \sum_{\bm{X}\in \Gamma^R_D} \left(\mathcal{N}_{T}-0\right)^2, \\
\mathcal{L}_{cnc} &= \dfrac{1}{N_{b}} \sum_{\bm{X}\in \Omega} \left( \left(\mathcal{N}_{q_{x}^o} - \mathcal{N}_{q_{x}}\right)^2  + 
\left(\mathcal{N}_{q_{y}^o} - \mathcal{N}_{q_{y}}\right)^2 \right), \\
\mathcal{L}_{SF} &= \dfrac{1}{N_{b}} \sum_{\bm{X}\in \Omega} \left(\partial_x \mathcal{N}_{q_{x}^o}\,+\, \partial_y \mathcal{N}_{q^o_{y}}\right)^2 ,\\
\mathcal{L}_{NBC} &= \sum_{\bm{X}\in \Gamma^T_N} \dfrac{1}{N^T_{NBC}}\left(\mathcal{N}_{q^o_{y}}-0\right)^2 \\
&+ \sum_{\bm{X}\in \Gamma^B_N}\dfrac{1}{N^B_{NBC}}\left(\mathcal{N}_{q^o_{y}}-0\right)^2
\end{align}
It is worth mentioning that the external work in Eq.~\ref{loss_EF_T} vanishes due to having $T=0$ on the right-hand side of the boundary value problem.
\\ \\ 
\textbf{Acknowledgements}:\\
The authors gratefully acknowledge the computing time granted on high performance computers of RWTH-Aachen university. 
\\ \\
\textbf{Author Statement}:\\
Shahed Rezaei: Conceptualization, Methodology, Supervision, Writing - Review \& Editing.
Ali Harandi: Methodology, Software, Writing - Review \& Editing.
Ahmad Moidendin: Methodology, Software, Writing - Review \& Editing.
Bai-Xiang Xu: Supervision, Review \& Editing.
Stefanie Reese: Supervision, Funding acquisition, Review \& Editing. 


\newpage
\bibliography{Ref}

\end{document}